\documentclass[12pt]{article}
\usepackage{amsfonts}
\usepackage{amsmath}
\usepackage{amsthm}
\usepackage{amssymb}
\usepackage{mathrsfs}
\usepackage{graphicx,psfrag,epsf}
\usepackage{enumerate}
\usepackage[round]{natbib}
\usepackage{xcolor}
\usepackage{float}
\usepackage{bbm}
\usepackage{mathtools}
\usepackage{subcaption}
\usepackage[colorlinks]{hyperref} 
\usepackage{changepage} 
\usepackage{multirow}
\usepackage{booktabs}
\usepackage{authblk}
\usepackage{blkarray}

\newtheorem{remark}{Remark}

\newtheorem{proposition}{Proposition}

\newtheorem{assumption}{Assumption}

\newtheorem{algorithm}{Algorithm}

\newcommand{\blind}{0}

\hypersetup{
	citecolor  = gray,
	linkcolor  = blue
}  

\newcommand*\dif{\mathop{}\!\mathrm{d}}

\DeclareMathOperator*{\argmin}{arg\,min}

\begin{document}
   \def\spacingset#1{\renewcommand{\baselinestretch}%
{#1}\small\normalsize} \spacingset{1}

    \if0\blind
    {
      \title{\bf A goodness-of-fit test for functional time series with applications to Ornstein-Uhlenbeck processes}
      \author[1]{J. \'Alvarez-Li\'ebana}
      \author[2]{A. L\'opez-P\'erez}
      \author[2]{W. Gonz\'alez-Manteiga}
      \author[2]{M. Febrero-Bande}
      \affil[1]{Department of Statistics and Data Science. Faculty of Statistics, Complutense University of Madrid}
      \affil[2]{Department of Statistics, Mathematical Analysis and Optimization. Universidade de Santiago de Compostela}
      \maketitle
    } \fi
	\if1\blind
    {
      \bigskip
      \bigskip
      \bigskip
      \begin{center}
        {\LARGE\bf A goodness-of-fit test for functional time series with applications to diffusion processes}
    \end{center}
      \medskip
    } \fi
	\begin{abstract}
		High-frequency financial data can be collected as a sequence of curves over time; for example, as intra—day price, currently one of the topics of greatest interest in finance. The Functional Data Analysis framework provides a suitable tool to extract the information contained in the shape of the daily paths, often unavailable from classical statistical methods. 
		In this paper, a novel goodness-of-fit test for autoregressive Hilbertian (ARH) models, with unknown and general order, is proposed. The test imposes just the Hilbert-Schmidt assumption on the functional form of the autocorrelation operator, and the test statistic is formulated in terms of a Cramér–von Mises norm. A wild bootstrap resampling procedure is used for calibration, such that the finite sample behavior of the test, regarding power and size, is checked via a simulation study. Furthermore, we also provide a new specification test for diffusion models, such as Ornstein-Uhlenbeck processes, illustrated with an application to intra-day currency exchange rates. In particular, a two-stage methodology is proffered: firstly, we check if functional samples and their past values are related via ARH(1) model; secondly, under linearity, we perform a functional F-test.
	\end{abstract}
	\noindent%
	{\it Keywords:} Currency exchange rates; Diffusion models; Functional time series; Goodness-of-fit; Specification test; Ornstein-Uhlenbeck process.
	
    \vfill
\newpage
\spacingset{1.45} 
\section{Introduction}
\label{sec:intro}

	Motivated by the recent availability of high-frequency data in finance, we here provide a twofold contribution in the field of temporally correlated functional data and diffusion processes. On the one hand, up to our knowledge, no proposals of Goodness-of-Fit (GoF) tests for autoregressive Hilbertian (ARH) processes, for unknown order, and against unspecified alternatives, can be found. On the other hand, few papers were devoted to specification tests in finance into a Functional Data Analysis setup. The main challenge and novelty of this work lies in covering both gaps. Regarding the former, we develop a composite null hypothesis test for functional time series in terms of a Cramér–von Mises statistic, such that no assumptions concerning alternatives are required. The latter is addressed by a characterization of Ornstein-Uhlenbeck (OU) processes as ARH models, providing a novel specification test in a high-dimensional setup, which could be extended to several diffusion processes, as long as they might be characterized as temporally correlated functional data.

	Modeling the relation between functional random variables (frv's) is one of the main topics in this context, where the Functional Linear Model with Functional Response (FLMFR) $\mathcal{Y} = m(\mathcal{X}) + \mathcal{E}$, with $m$ a linear operator and $\mathcal{E}$ a functional error, is likely the best-known parametric model. Within a Hilbertian framework, the operator $m$ is typically assumed to be a Hilbert-Schmidt integral operator between $L^2$ spaces. 
	The Goodness-of-Fit setup for regression models is a mighty field in which several authors have contributed, most of them keeping integrated-regression-based methodologies  proposed by \cite{Stute97}. This framework was extended to functional setups in \cite{GarciaPortuguesetal14}, \cite{CuestaAlbertosetal19} and \cite{GarciaPortuguesetal20}.
	As a particular case of linear models, functional time series were widely studied, mainly focusing on ARH processes, in which functional regressors are given by their own past values. Early results on the parametric moment-based estimation were proposed by \cite{Bosq00}, on the estimation of ARH(1) models $\mathcal{X}_{n} (\cdot) = \Gamma \left(\mathcal{X}_{n-1} \right) \left(\cdot \right) + \mathcal{E}_n \left(\cdot \right)$, with $n \in \mathbb{Z}$. Recently, \cite{AlvarezLiebanaetal17} proposed a plug-in FPCA-based estimator, and \cite{RuizMedinaAlvarezLiebana19c} exploits the Hilbert space structure with an extension to Banach space context.
	
	Notwithstanding the growing interest in functional time series, up to our knowledge, no proposals of GoF tests in the ARH($z$) setup, with unspecified alternatives, can be found in the literature. This lack of approaches probably reflects the fact that, besides the GoF test recently proposed in \cite{GarciaPortuguesetal20}, there are no proposals of GoF tests in the FLMFR setup neither, against unspecified alternatives (see \citealp{Kokoszkaetal08} and \citealp{Patileaetal16} about testing no effects hypothesis). One of the most crucial aspects working with functional time series is checking whether just a single model can be used, not rupturing the underlying stochastic structure. As early contributions, \cite{LaukaitisRackauskas02} analyzed the functional residuals for change-point detection, whereas \cite{Berkesetal09} focused on testing the assumption of a common functional mean of identically and independent distributed (iid) samples. We also refer to \cite{GabrysKokoszka07}, for testing iid frv's. \cite{Horvathetal10} proposed a methodology on checking if the autocorrelation operator involved remains unchanged with time. Another aspect to ensure meaningful inference relies in testing the assumption of stationarity, within a functional data framework. With this purpose, KPSS tests were formalized in \cite{Horvathetal14}. 
	
	In this work, we provide a GoF test for ARH($z$) processes, under the null hypothesis
	$$\mathcal{H}_0\colon~\mathcal{X}_n \text{ and } \left(\mathcal{X}_{n-1}, \ldots, \mathcal{X}_{n-z} \right) \text{ linearly related}, \quad (\text{i.e., }\mathcal{X}_{n} (t) = \sum_{r=1}^{z} \Gamma_r (\mathcal{X}_{n-r})(t) + \mathcal{E}_n (t)),$$
	against an unspecified alternative hypothesis, where $\left\lbrace \Gamma_r \right\rbrace_{r=1}^{z}$ is a set of linear operators $\Gamma_{\rho}$ given by $\Gamma_{\rho} \left(\mathcal{X} \right)(t) = \int_{0}^{h} \rho(s,t) \mathcal{X}(s) \dif s$, with $ \int_{0}^{h}\int_{0}^{h} \rho^2(s,t) \dif s \dif t < \infty$ and $z \geq 1$. The methodology here proposed is based on reinterpreting an ARH($z$) process as a particular FLMFR and the characterization of $\mathcal{H}_0$ in terms of the integral regression operator derived, projected into finite-dimensional functional directions, in keeping with the work by \cite{GarciaPortuguesetal20}. The deviation of data from $\mathcal{H}_0$ is measured by a Cramér–von Mises norm, and the resulting statistic is calibrated via wild bootstrap. The novelty of contribution is twofold: i) a new GoF test for ARH($z$) processes, given an integer and positive order $z \geq 1$; ii) a sequential procedure to determine the order of an ARH model. As the former contribution has no competitors, we compare the latter procedure against the order detection procedure in \cite{KokoszkaReimherr12}. Note that their methodology was developed just considering ARH alternatives,  in contrast with our unspecified one.
	
	Due to the key role of diffusion processes in finance, we also seek to develop a framework for model specification of diffusion models from a functional perspective.
	Regarding GoF tests, there exist several proposals for univariate continuous–time models, some of them in terms of the marginal density function of the process (\citealp{AitSahalia96} and \citealp{GaoKing04}); the transitional density (\citealp{HongLi04}; \citealp{Chenetal08}); the cumulative distribution function \citep{Corradi2005}; the conditional characteristic function \citep{Chen2010}; or the infinitesimal operator \citep{Song2011}. Nonparametric techniques were alternatively proposed by \cite{ArapisGao06}, \cite{GaoCasas08} or \cite{Zheng2009}; whilst \cite{FanZhang03}, \cite{Fanetal03} and \cite{AitSahalia2009} proposals were based on likelihood ratio test ideas and \cite{Chen2011} on empirical likelihood. \cite{Dette2003}, \cite{Dette2008} and \cite{Podolskij2008} used stochastic processes based on the integrated volatility function and \cite{Negri2009}, \cite{MonsalveCobisetal11} and \cite{Chen2015} tests were based on marked empirical processes.
	
	In this classical context, diffusion models for financial data are commonly used at daily, weekly or monthly frequency. However, high-frequency data constitutes time series at a very fine resolution, and using daily data would imply discarding large fractions of observations. It is in this high frequency scenario where dealing with diffusion models using a functional data approach allows to model the realized daily trajectories, analyzing the patterns of the curves across the days. Since Vasicek model was used to capture the dynamics of interest rates, the OU process \citep{UhlenbeckOrnstein30} is one of the main diffusion models. In this paper, as an alternative for multivariate settings, we propose a novel two-stage specification test for OU processes: in the first step, we verify whether the underlying stochastic differential equation can be characterized via ARH(1) process, based on an ARH(1) characterization; secondly, under linearity, we implement a functional F-test.
	
	The rest of this article is organized as follows.  In Section~\ref{sec:fda}, a brief introduction to functional data, FLMFR and ARH($z$) models, is provided. A GoF test for the referred autoregressive models is detailed in Section~\ref{sec:GoF}. A simulation study is undertaken in Section~\ref{sec:sim_ARHp}, comparing it with the proposal in \cite{KokoszkaReimherr12}. Section~\ref{sec:sde} proposes a new specification test for diffusion models. The performance of the specification test is also depicted by means of a wide simulation study. In Section~\ref{sec:real_data}, a real-data application to daily currency exchange rates curves is addressed. Section~\ref{sec:conc} concludes the paper and theoretical details are relegated to the Supplementary Material.
\section{Background: FLMFR and functional time series}
\label{sec:fda}

    %
    %
    %
    Whilst Hilbert spaces are the common option, it is well worth canvassing the alternatives which could be adopted. The most general context may be given by a metric space endowed with a distance. In contrast with metric spaces, useful when no information about curves are available but rather abstract since a norm cannot be guaranteed, a Banach space $\left(\mathbb{B}, \left\| \cdot \right\|_{\mathbb{B}}\right)$ may be adopted in place, say $\mathcal{C}([0,1])$. Unless otherwise explicitly mentioned, we consider separable Hilbert spaces $\left(\mathbb{H}, \langle \cdot, \cdot \rangle_{\mathbb{H}} \right)$, where $\langle \cdot, \cdot \rangle_{\mathbb{H}}$ denotes an inner product. This common choice is not arbitrary since, under separability, the existence of a countable functional basis  being is guaranteed. In what follows, $\left\lbrace \Psi_j \right\rbrace_{j=1}^{\infty}$ and $\left\lbrace \Phi_k \right\rbrace_{k=1}^{\infty}$ are orthonormal functional bases. From separability, any $\mathcal{X} \in \mathbb{H}_1$ and $\mathcal{Y} \in \mathbb{H}_2$ can be represented as $\mathcal{X} = \sum_{j=1}^{\infty} x_j \Psi_j$ and $\mathcal{Y} = \sum_{k=1}^{\infty} y_k \Phi_k$, with $x_j = \langle \mathcal{X}, \Psi_j \rangle_{\mathbb{H}_1}$ and  $y_k = \langle \mathcal{Y}, \Phi_k \rangle_{\mathbb{H}_2}$, for each $j,k\geq 1$.
        
        Since sparsity effects usually appear, dimension reduction is here essential. This sparsity can be understood in a wide sense, such as referring to the sparsity of the model \citep{AneirosVieu14}, the sparsity of data \citep{Vieu18} or related to the discretization grid \citep{Yaoetal05}. We can further distinguish among fixed bases (e.g., B-splines), flexible but usually require a larger number of elements, and data-driven functional bases. More parsimonious representations can be achieved with the latter ones, being the most popular choice the (empirical) Functional Principal Components (FPC), given as the eigenfunctions of the empirical covariance operator. $C_{n}^{\mathcal{X}}  = \frac{1}{n} \sum_{i=1}^{n} \mathcal{X}_i \otimes \mathcal{X}_i$. Henceforth, for given finite cut-off levels $p,q \geq 1$, we define the $(p,q)$-truncated bases expansions as $\mathcal{X}^{(p)} = \sum_{j=1}^{p} x_j \Psi_j$ and $\mathcal{Y}^{(q)} = \sum_{k=1}^{q} y_k \Phi_k$, with $\left(x_1, \ldots, x_p \right) \in \mathbb{R}^p$ and $\left(y_1, \ldots, y_q \right) \in \mathbb{R}^q$, such that $C_{n}^{\mathcal{X}} \left( \Psi_j  \right)= \lambda_{j}^{\Psi} \Psi_j$ and $C_{n}^{\mathcal{Y}} \left( \Phi_k \right) = \lambda_{k}^{\Phi} \Phi_k$, for each $j=1,\ldots,p$ and $k = 1,\ldots,q$.
    \subsection{The FLMFR}
    \label{sec:flmfr}

        In this work, we will assume that $\mathcal{X} \in \mathbb{H}_1 = L^2 ([a,b])$, $\mathcal{Y} \in \mathbb{H}_2 =  L^2 ([c,d])$  are centered frv`s, such that $\langle f,g \rangle_{L^2 ([a,b])} = \int_{a}^{b} f(t) g(t) \dif t$. We will characterize a functional time series as a particular linear model, and so, the following FLMFR setting will be introduced:
    	\begin{equation} \label{eq_1}
    	\mathcal{Y} = m_{\beta} \left(\mathcal{X} \right) + \mathcal{E}, \qquad {\rm E} \left[\mathcal{E} | \mathcal{X} \right] = 0, \qquad m_{\beta}(x) = {\rm E} \left[ \mathcal{Y} | \mathcal{X} = x  \right]~\text{Hilbert-Schmidt},
    	\end{equation}
    	that is, $m_{\beta}$ admits an integral representation given by a bivariate and square-integrable kernel $\beta \in \mathbb{H}_1 \otimes \mathbb{H}_2$. In the aftermath of the compactness, $\beta$ can be decomposed as follows:
    	\begin{equation}
    	m_{\beta} \left(\mathcal{X} \right) (\cdot) = \int_{a}^{b} \beta(s, \cdot) \mathcal{X} (s) \dif s, \qquad \beta  = \sum_{j=1}^{\infty} \sum_{k=1}^{\infty} \beta_{jk} \left( \Psi_j \otimes \Phi_k\right), \qquad \int \int \beta^2 (s,t)  < \infty, \nonumber
    	\end{equation}
    	where $\beta_{jk} = \langle \beta, \Psi_j \otimes \Phi_k \rangle_{\mathbb{H}_1 \otimes \mathbb{H}_2}$, for each $j,k \geq 1$. Adopting a $(p, q)$-truncated component-wise orthonormal expansion, projected into  $\left\lbrace \Psi_j \right\rbrace_{j=1}^{p}$, $\left\lbrace \Phi_k \right\rbrace_{k=1}^{q}$ and $\left\lbrace \Psi_j \otimes \Phi_k \right\rbrace_{j,k=1}^{p,q}$,  we  get
    	\begin{equation} \label{eq_2}
    	\mathcal{Y}^{(q)} = \sum_{k=1}^{q} y_k \Phi_k, \quad y_k = \sum_{j=1}^{p} \sum_{\ell=1}^{p} b_{\ell,k} x_j \langle \Psi_j , \Psi_{\ell} \rangle_{\mathbb{H}_1} + e_k,~ k = 1,\ldots, q, \quad  \mathcal{E}^{(q)} = \sum_{k=1}^{q} e_k \Phi_k.
    	\end{equation}
    	Now, given a centered sample $\left\lbrace \left(\mathcal{X}_i, \mathcal{Y}_i \right) \right\rbrace_{i=1}^{n}$, equations \eqref{eq_1}--\eqref{eq_2} may be expressed as:
    	\begin{equation} \label{eq_3}
    	\textbf{Y}_q = \textbf{X}_p \textbf{B}_{p,q} + \textbf{E}_q, \quad \mathcal{Y}_i = \langle \langle \mathcal{X}_i, \beta \rangle\rangle + \mathcal{E}_i, \quad \langle \langle \mathcal{X}, \beta \rangle\rangle(t) \coloneqq \langle \mathcal{X}, \beta(\cdot, t) \rangle_{\mathbb{H}_1}, 
    	\end{equation}
    	where $\textbf{Y}_q$ and $\textbf{E}_q$ are the $n\times q$ matrices with the coefficients of $\left\lbrace  \mathcal{Y}_i \right\rbrace_{i=1}^{n}$ and $\left\lbrace  \mathcal{E}_i \right\rbrace_{i=1}^{n}$, respectively, on $\left\lbrace \Phi_k \right\rbrace_{k=1}^{q}$,  $\textbf{X}_p$ is the $n\times p$ matrix with the coefficients  of $\left\lbrace  \mathcal{X}_i \right\rbrace_{i=1}^{n}$ on $\left\lbrace \Psi_j \right\rbrace_{j=1}^{p}$ and $\textbf{B}_{p,q}$ is the $p\times q$ matrix with the coefficients (to be estimated) of $\beta$ on $\left\lbrace \Psi_j \otimes \Phi_k \right\rbrace_{j,k=1}^{p,q}$.
    	
    	From now on, we consider the hybrid linearly-constrained FPC Regression (FPCR-L1S) estimator recently formulated in \cite{GarciaPortuguesetal20}, defined as follows:
    	\begin{equation} \label{eq_4}
    	\hat{\textbf{Y}}_q = \widetilde{\textbf{X}}_{\widetilde{p}} \hat{\textbf{B}}_{\widetilde{p},q}^{(\lambda), C} = \textbf{H}_{C}^{(\lambda)} \textbf{Y}_q = \left[ \widetilde{\textbf{X}}_{\widetilde{p}} \left(\widetilde{\textbf{X}}_{\widetilde{p}}^{T} \widetilde{\textbf{X}}_{\widetilde{p}} \right)^{-1} \widetilde{\textbf{X}}_{\widetilde{p}}^{T} \right] \textbf{Y}_q, \quad \hat{\textbf{B}}_{\widetilde{p},q}^{(\lambda), C} = \left(\widetilde{\textbf{X}}_{\widetilde{p}}^{T} \widetilde{\textbf{X}}_{\widetilde{p}} \right)^{-1} \widetilde{\textbf{X}}_{\widetilde{p}}^{T} \textbf{Y}_q.
    	\end{equation}
    	The estimated $\hat{\textbf{Y}}_q $  in \eqref{eq_4} are computed in four steps: \textit{(i)} orthonormal bases $\left\lbrace \Psi_j \right\rbrace_{j=1}^{p}$ and $\left\lbrace \Phi_k \right\rbrace_{k=1}^{q}$ are defined as the (empirical) FPC of $\mathcal{X}$ and $\mathcal{Y}$, respectively (in what follows, $\left\lbrace \Psi_j \right\rbrace_{j=1}^{p}$ and $\left\lbrace \Phi_k \right\rbrace_{k=1}^{q}$ are the eigenfunctions of the empirical covariance operators $C_{n}^{\mathcal{X}}$ and $C_{n}^{\mathcal{Y}}$, respectively); \textit{(ii)} initial values $(p, q)$ are fixed with regard to a certain proportion of Explained Variance $\text{EV}_p$ and $\text{EV}_q$ (say $\text{EV}_p = \text{EV}_q \geq 0.99$), where $\text{EV}_p = \sum_{j=1}^{p} \lambda_{j}^{\Psi}$ and $\text{EV}_q = \sum_{k=1}^{q} \lambda_{k}^{\Phi}$, being $\left\lbrace \lambda_{j}^{\Psi} \right\rbrace_{j=1}^{p}$ and $\left\lbrace \lambda_{k}^{\Phi} \right\rbrace_{k=1}^{q}$ the associated eigenvalues; \textit{(iii)} we implement a variable selection, based on a LASSO (L1) regularization, by considering the no null rows of $\hat{\textbf{B}}_{p,q}^{(\lambda)} =  \argmin\limits_{\textbf{B}_{p,q}} \left\lbrace \frac{1}{2n} \sum_{i=1}^{n} \left\| \left(\textbf{Y}_q \right)_i \left(\textbf{X}_p \textbf{B}_{p,q} \right)_i \right\|^2 + \lambda \sum_{j=1}^{p} \left\| \left(\textbf{B}_{p,q} \right)_j \right\|_2 \right\rbrace$, being $\left(\textbf{Y}_q \right)_i$  the $i$-th row of $\textbf{Y}_q$; \textit{(iv)} we perform a FPCR estimation just using the set of $\widetilde{p} \leq p$ predictors selected, 
    	whose coefficients are denoted as $\widetilde{\textbf{X}}_{\widetilde{p}}$, and therefore, having a hat matrix $\textbf{H}_{C}^{(\lambda)}$ at our disposal, which will be crucial within the bootstrap algorithm.

    	Hence, the FPCR-L1-selected (FPCR-L1S) estimator in \eqref{eq_4} allows us to seize the advantages of both estimation paradigms. This estimator critically depends on the penalty parameter $\lambda$. Whilst $\hat{\lambda}_{CV}$ (leave-one-out cross-validation) is an optimal choice for estimating $\beta$, we adopt the so-called one standard error rule (see \citealp{Friedmanetal10}), denoted in the GoF folklore as $\hat{\lambda}_{1SE}$ (a variance reduction is achieved to obtain a more biased estimator for test calibration). Both approaches were implemented in the R package \textsf{goffda} \citep{GarciaPortuguesAlvarezLiebana19}. Estimator in \eqref{eq_4} could be extended to nonorthonormal bases, just replacing $\textbf{X}_{p}$ by $\textbf{\u{X}}_{p} =\textbf{X}_{p} \boldsymbol{\Psi}$ in \eqref{eq_3}, where $\boldsymbol{\Psi}= \left(\langle \Psi_j, \Psi_j \rangle_{\mathbb{H}_1} \right)_{i,j=1,\ldots,p}$
    \subsection{Functional time series: autoregressive Hilbertian processes}
	\label{sec:FTS}
	
\begin{figure}[H]
		\centering
		\begin{subfigure}{.47\textwidth}
			\centering
			\includegraphics[width=1\linewidth]{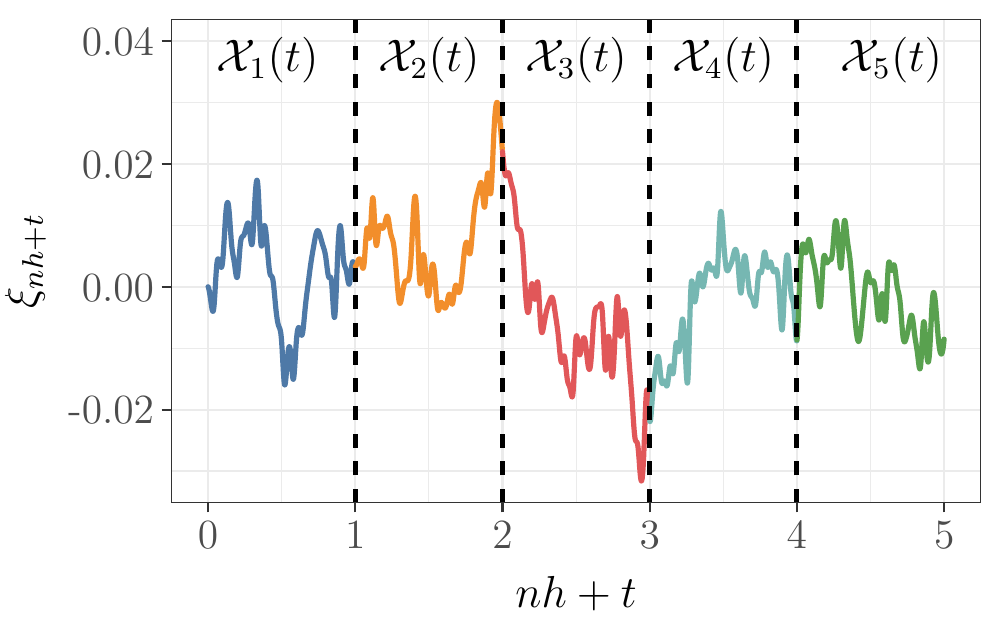}
			\captionsetup{font=footnotesize,textfont=it}
			\caption{Continuous path of an OU process $\left\lbrace \xi_t \right\rbrace_{t \in \mathbb{R}^{+}}$, splitted into $[0,h]$ ($h=1$), such that $\mathcal{X}_i(t) = \xi_{ih + t}$.}
			\label{fig:SDEpath}
		\end{subfigure}
		\quad
		\begin{subfigure}{.47\textwidth}
			\centering
			\includegraphics[width=1\linewidth]{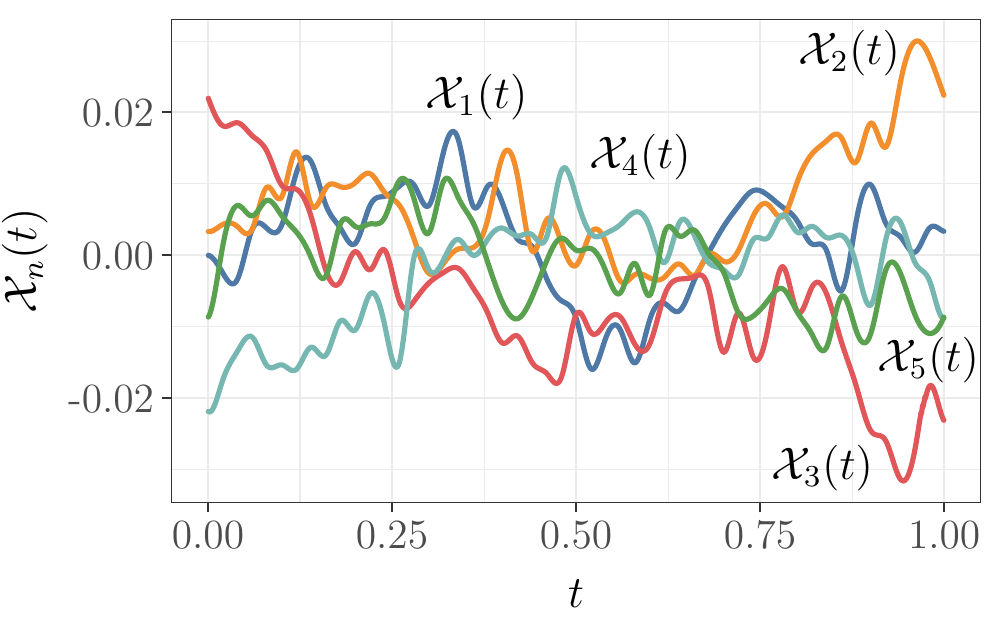}
			\captionsetup{font=footnotesize,textfont=it}
			\caption{OU process characterized as a set of ARH(1) trajectories $\left\lbrace \mathcal{X}_n(t)\colon t \in [0,1] \right\rbrace_{i=1,\ldots,n}$, with $n = 5$.}
			\label{fig:FDpath}
		\end{subfigure}
		\caption{Stochastic process $\left\lbrace \xi_t \right\rbrace_{t \in \mathbb{R}^{+}}$ characterized and splitted as an ARH(1) process.}
		\label{fig:1}
	\end{figure}

	Since the idea underlying the methodology proposed in Section~\ref{sec:GoF} revolves around characterizing ARH processes as a particular case of FLMFR, firstly let us see some preliminary elements on these models from a time-domain perspective. Given a probability space $\left(\Omega, \mathcal{A}, \mathbb{P} \right)$, let $\left\lbrace \xi_t \right\rbrace_{t \in \mathbb{R}}$ be now a continuous-time zero-mean stochastic process. In keeping with \cite{Bosq00}, we split the paths as $ \mathcal{X}_n(t) = \xi_{nh + t}$, with $ t \in \left[0, h \right]$, and $\mathcal{X}_n \in \mathbb{H} = L^2\left(\left[0, h \right]\right)$, for each  $n \in \mathbb{Z}$, constituting an infinite-dimensional discrete-time process. This representation (see Figure~\ref{fig:1}) is especially fruitful if $\left\lbrace \xi_t \right\rbrace_{t \in \mathbb{R}}$ displays  a seasonal component either it will be forecasted over $[0,h]$.
    We say that  $\mathcal{X} = \left\lbrace \mathcal{X}_n \right\rbrace_{n \in \mathbb{Z}}$  is a zero-mean autoregressive Hilbertian process of order one, valued in $\mathbb{H} = L^2([0,h])$, denoted as ARH(1), if the following state equation is satisfied
    \begin{equation}
    	\mathcal{X}_n (t) = \Gamma \left(\mathcal{X}_{n-1} \right)(t) + \mathcal{E}_n(t), \quad n \in \mathbb{Z}, \quad  \mathcal{X}_n,~ \mathcal{E}_n \in \mathbb{H}=L^2([0,h]), \quad t \in [0,h], \label{eq_6}
	\end{equation}
	where $\Gamma \equiv \Gamma_{\rho}$ denotes the linear autocorrelation operator, with $\left\| \Gamma \right\|_{\mathcal{L}(\mathbb{H})} = \displaystyle \sup_{\left\| \mathcal{X} \right\|_{\mathbb{H}} \leq 1} \left\| \Gamma \left(\mathcal{X} \right) \right\|_{\mathbb{H}}$, and $\left\lbrace \mathcal{E}_n \right\rbrace_{n \in \mathbb{Z}}$ is assumed to be a strong-white noise, with $\sigma_{\mathcal{E}}^{2} = {\rm E} \left[ \left\| \mathcal{E}_n \right\|_{\mathbb{H}}^2 \right] < \infty$ and iid components. The subsequent assumptions will be considered:
    	
	\medskip
	
	\begin{assumption}
	\label{as:1}
    	$\left\| \Gamma^k \right\|_{\mathcal{L}(\mathbb{H})} < 1$, for any $k \geq k_0$, and for some $k_0 \geq 1$, where $\Gamma^k$  denotes the composition operator $\Gamma \overbrace{\ldots}^{k} \Gamma$, leading to $\sum_{n=0}^{\infty} \left\| \Gamma^n \right\|_{\mathcal{L}\left( \mathbb{H} \right)} < \infty$.
    \end{assumption}
    \begin{assumption}
	\label{as:2}
    	The autocorrelation  operator is given by $\Gamma_{\rho} \left( \mathcal{X} \right) (t) = \displaystyle \int_{0}^{h} \rho(s,t) \mathcal{X}(s) \dif s$, with $\rho = \sum_{j=1}^{\infty} \sum_{k=1}^{\infty} \rho_{jk} \Psi_j \otimes \Psi_k$, such that $\int \int  \rho^2(s,t) \dif s \dif t < \infty$ (Hilbert–Schmidt integral operator) and $\rho_{jk} = \langle \rho \left( \Psi_j \right), \Psi_k \rangle_{\mathbb{H}}$, for each $j,k = 1,\ldots,\infty$.
    \end{assumption}

		Remark that the autocovariance operator $C \coloneqq {\rm E} [\mathcal{X}_n \otimes \mathcal{X}_n] $ is a self-adjoint, trace and positive operator, for each $n \in \mathbb{Z}$. As a result, it admits a diagonal FPC-based decomposition $C = \sum_{j=1}^{\infty}  C_j \Psi_j \otimes \Psi_j$, being $\left\lbrace \Psi_j \right\rbrace_{j=1}^{\infty}$ the theoretical eigenfunctions of $C$, associated with eigenvalues $C_1 \geq \ldots \geq C_j \geq \ldots > 0$. From \citet[Theorem 3.1]{Bosq00}, \textbf{Assumption~\ref{as:1}} is required for ensuring an unique stationary solution. Since $C^{-1} =  \sum_{j=1}^{\infty}  \frac{1}{C_j} \Psi_j \otimes \Psi_j $, under the trace property, $C$ cannot be inverted, and thus, $\Gamma \coloneqq D C^{-1}$ should be estimated. Note that  \textbf{Assumption~\ref{as:2}} does not imply that $D C^{-1}$  could be diagonally decomposed, since the symmetry of the cross-covariance operator $D \coloneqq {\rm E} \left[\mathcal{X}_n \otimes \mathcal{X}_{n+1} \right]$ is not imposed. 
	%
	\begin{remark}
	    As a sideways contribution, we provide in the Supplementary Material a strongly consistent estimator of $\Gamma$, under weaker conditions that those ones in \cite{Bosq00}, improving the decay rate of convergence of the associated predictor. The estimation under the compactness of $D$ was achieved in \cite{AlvarezLiebanaetal17}. 
    \end{remark}
\section{A GoF test for functional autoregressive processes}
	\label{sec:GoF}
    We now propose a new GoF test for ARH processes against unspecified alternative. Firstly, we will extend the ARH(1) process in Section~\ref{sec:FTS} to ARH($z$) models, even with $z > 1$. Secondly, we will characterize these ARH($z$) processes as a particular case of FLMFR, under the setting established in Section~\ref{sec:flmfr}. Lastly, we will detail the GoF test proposal for ARH($z$) processes, from a FLMFR perspective, and implement a simulation study.
    
    \subsection{ARH(z) processes: state equation}
    \label{sec:arhz}
	The Markovianess of the ARH(1) model in \eqref{eq_6} allows us to easily generalized it  by including more lagged functional regressors, i.e., $\mathcal{X}_n$ is inferred from $\left(\mathcal{X}_{n-1}, \ldots, \mathcal{X}_{n-z}\right)$.
	Formally, we say that $\mathcal{X} = \left\lbrace \mathcal{X}_n \right\rbrace_{n \in \mathbb{Z}}$ is a zero-mean autoregressive Hilbertian process of order $z \geq 1$, valued in $\mathbb{H} = L^2([0,h])$ and denoted as ARH($z$), if the following state equation is satisfied:
	\begin{equation}
	\mathcal{X}_n(t) = \sum_{r=1}^{z} \Gamma_r \left(\mathcal{X}_{n-r} \right)(t) + \mathcal{E}_n(t), \quad z \geq 1,  \quad \mathcal{X}_n, \mathcal{E}_n \in \mathbb{H}, \quad t \in [0,h], \quad n \in \mathbb{Z},\label{eq_8}
	\end{equation}
	where $\left\lbrace \Gamma_r \right\rbrace_{r=1}^{z}$ are bounded linear operators in $\mathcal{L}\left(\mathbb{H} \right)$, and $\left\lbrace \mathcal{E}_n \right\rbrace_{n \in \mathbb{Z}}$ is a $\mathbb{H}$-valued strong white noise, such that $\mathbb{P} \left( \Gamma_z \left( \mathcal{X}_n \right) \neq 0 \right) > 0$ is implicitly assumed, for each $n \in \mathbb{Z}$. 
	ARH(1) estimation results can be  effortlessly extended to this context since ARH($z$) process in \eqref{eq_8} can be reinterpreted a particular multivariate ARH(1) process, valued in  $\mathbb{H}^z \coloneqq \prod_{r=1}^{z} \mathbb{H}$, constituting likewise a separable Hilbert space (see Lemma 1 in the Supplementary Material). 
	In this way, the following proposition allows us to characterize the ARH($z$) in \eqref{eq_8} as a $\mathbb{H}^z$-valued stationary ARH(1) process. 
	See proofs 
	in the Supplementary Material.
	
	\medskip
	
	\begin{proposition}
		\label{prop:1}
		Let $\mathcal{X} \coloneqq \left\lbrace \mathcal{X}_n \right\rbrace_{n \in \mathbb{Z}}$ be a zero-mean ARH(z) process valued in $\mathbb{H} = L^2([0,h])$, with $z \geq 1$. The ARH(z) model in  \eqref{eq_8} can be reinterpreted as
		\begin{equation}
		\overline{\mathcal{X}}_n = \begin{pmatrix} \mathcal{X}_n \\
		\vdots \\
		\mathcal{X}_{n-z+1} 
		\end{pmatrix} \in \mathbb{H}^{z}, \quad \overline{\Gamma} = \begin{pmatrix}
		\Gamma_1 &  \ldots & \Gamma_{z-1} & \Gamma_z\\
		Id_{\mathbb{H}} & \ldots & 0_{\mathbb{H}} & 0_{\mathbb{H}}\\
		\vdots & \ddots & \vdots & \vdots\\
		0_{\mathbb{H}} & \ldots & Id_{\mathbb{H}} & 0_{\mathbb{H}}
		\end{pmatrix}, \quad \overline{\mathcal{E}}_n = \left(\mathcal{E}_n, \mathbf{0}, \ldots, \mathbf{0}\right)^{T} \in \mathbb{H}^{z}, \label{eq_9}
		\end{equation}
		where $\mathbb{H}^z$ is the Cartesian product of $z$  copies of $\mathbb{H}$. Thus, $\overline{\mathcal{X}}_n  =  \overline{\Gamma} \left(\overline{\mathcal{X}}_{n-1} \right) + \overline{\mathcal{E}}_n$, constitutes a $\mathbb{H}^z$-valued ARH(1) process, for each $n \in \mathbb{Z}$, with $\overline{\mathcal{E}}_n$ a $\mathbb{H}^z$-valued strong white noise. Furthermore, the joint operator $\overline{\Gamma}$ is also a bounded linear operator in $\mathbb{H}^z$. In \eqref{eq_9}, $Id_{\mathbb{H}}$ and $0_{\mathbb{H}}$ denote the identity and null operators, respectively, and $ \mathbf{0}$ the null element on $\mathbb{H}$. 
	\end{proposition}
	
	\medskip
	
	Concerning \textbf{Assumption~\ref{as:1}}, a sufficient condition for the existence of an unique stationary solution of  \eqref{eq_8} is  provided in Proposition \ref{prop:2} below.
	
	\medskip
	
	\begin{proposition}
		\label{prop:2}
		Let $\mathcal{X} = \left\lbrace \mathcal{X}_n \right\rbrace_{n \in \mathbb{Z}}$ be a zero-mean ARH(z) process, with $z \geq 1$, as explicitly defined in \eqref{eq_8}--\eqref{eq_9}. If $\left\| \overline{\Gamma}^k \right\|_{\mathcal{L}(\mathbb{H}^z)} < 1$, for any $k \geq k_0$, and for some $k_0 \geq 1$, 
		then the state equation established in  \eqref{eq_8} has a unique stationary solution given by $\mathcal{X}_n = \sum_{j=0}^{\infty} \Pi_1 \left(\overline{\Gamma}^{j}  \left(\overline{\mathcal{E}}_{n-j} \right) \right)$, for each $n \in \mathbb{Z}$, where $\Pi_r\colon\left(\mathcal{X}_{(1)}, \ldots, \mathcal{X}_{(z)}\right) \mapsto \mathcal{X}_{(r)}$, with $r=1,\ldots,z$.
	\end{proposition}

	\begin{remark}
	\label{rem:2}
		As proved in Lemma 2 in the Supplementary Material, \textbf{Assumption~\ref{as:1}} on $\overline{\Gamma}$ also implies that  \textbf{Assumption~\ref{as:1}} is held for each $\Gamma_r$, with $r=1,\ldots,z$ (i.e., stationarity condition  in Proposition \ref{prop:2} leads to $\sum_{n=0}^{\infty} \left\| \Gamma_{r}^n \right\|_{\mathcal{L}\left( \mathbb{H} \right)} < \infty$). Remark also that, from the definition of  norm $\left\| \cdot \right\|_{\mathbb{H}^z}$,  \textbf{Assumption~\ref{as:2}} on $\overline{\Gamma}$  is verified as long as \textbf{Assumption~\ref{as:2}} is satisfied for  each $\Gamma_r$, for any $r=1,\ldots,z$. 
		Note that now $\overline{C} \coloneqq {\rm E} \left[  \overline{\mathcal{X}}_n \otimes_{\mathbb{H}^z}  \overline{\mathcal{X}}_n \right] $ is the autocovariance  operator, where $\otimes_{\mathbb{H}^z} $ denotes the tensor product on $\mathbb{H}^z$, given by
		\begin{equation}
		\overline{C} (\overline{\mathcal{X}})= \sum_{j=1}^{\infty} \overline{C}_j \overline{\Psi}_j \left\langle \overline{\Psi}_j, \overline{\mathcal{X}}  \right\rangle_{\mathbb{H}^z}, \quad \left\langle \overline{C} (\overline{\mathcal{X}}),  \overline{\mathcal{Y}} \right\rangle_{\mathbb{H}^z} = \sum_{j=1}^{\infty} \overline{C}_j \left\langle \overline{\Psi}_j, \overline{\mathcal{X}} \right\rangle_{\mathbb{H}^z} \left\langle \overline{\Psi}_j, \overline{\mathcal{Y}} \right\rangle_{\mathbb{H}^z}, \quad \overline{\mathcal{X}},~\overline{\mathcal{Y}} \in \mathbb{H}^z, \nonumber 
		\end{equation}
		where $\left\lbrace \overline{\Psi}_j \right\rbrace_{j=1}^{\infty}$ and $\left\lbrace \overline{C}_j \right\rbrace_{j=1}^{\infty}$ denote their $\mathbb{H}^z$-valued FPC and their eigenvalues, respectively. The joint  operator $\overline{C}$ can be re-interpreted as $ \overline{C} = \left( C_{r,s} \right)_{r=1,\ldots,z}^{s=1,\ldots,z}$, where $C_{r,s} ={\rm E} \left[ \mathcal{X}_{n, (r)} \otimes \mathcal{X}_{n, (s)} \right]$, being $ \mathcal{X}_{n, (r)}$ the $r$-th element of $\overline{\mathcal{X}}_n$, for each $n \in \mathbb{Z}$ and $r=1,\ldots,s$. 
	\end{remark}
	
    \subsection{ARH($z$) processes characterized as FLMFR}
	\label{sec:arh_as_flmfr}
	
    	Notwithstanding a strongly consistent estimator of $\Gamma$ in the ARH setting is provided in the Supplementary Material, its expression may be intricate, and the lack of an explicit hat matrix implies a considerably increase in the computational cost. For these reasons, an alternative characterization of the ARH(z) model in \eqref{eq_8}, based on properties detailed in Section~\ref{sec:arhz}, is now suggested, re-expressed it as a FLMFR. As commented, from Remark~\ref{rem:2}, under \textbf{Assumption~\ref{as:2}} on each $\left\lbrace \Gamma_r \right\rbrace_{r=1}^{z}$, we have
    	\begin{equation} \label{eq_FLMFR_ARH}
        	\Gamma_r \left(\mathcal{X} \right)(t) = \displaystyle \int_{0}^{h} \rho_{r} (s,t) \mathcal{X}(s) \dif s, \quad  \rho_{r} = \sum_{j=1}^{\infty}\sum_{k=1}^{\infty} \rho_{jk}^{(r)} \Psi_j \otimes \Psi_k, \quad  \displaystyle \int_{0}^{h}  \displaystyle \int_{0}^{h}  \rho_{r}^2 (s,t) \dif s \dif t < \infty. 
    	\end{equation}
    	Given a zero-mean ARH(z) process in \eqref{eq_8}, we define $\widetilde{\mathcal{X}}_n(s) = \sum_{r=1}^{z} \mathcal{X}_{n-r} \left(s  z - \left(r - 1 \right) \right) \mathbbm{1}_{r}(s)$, for each $s \in \left[0,  h \right]$, where $\mathbbm{1}_{r}(\cdot) \coloneqq \mathbbm{1}_{\left[\frac{(r-1)h}{z}, \frac{rh}{z} \right]}(\cdot)$ denotes the indicator function in  $\left[\frac{(r-1)h}{z}, \frac{rh}{z} \right]$, for each $r=1,\ldots,z$. 
    	In the same way, a change of variable $x \coloneqq \left(s + r - 1 \right)/z$ yields
    	\begin{equation}
    	\Gamma_r \left(\mathcal{X}_{n-r} \right)(t) = \displaystyle \int_{0}^{h} \rho_{r} (s,t) \mathcal{X}_{n-r}(s) \dif s = z \displaystyle \int_{\frac{(r-1)h}{z}}^{\frac{rh}{z}}  \rho_{r} (xz - (r-1),t) \mathcal{X}_{n-r}(xz - (r-1)) \dif x,\nonumber
    	\end{equation}
    	for each $r=1,\ldots, z$ and $t \in [0,h]$, and then,
    	\begin{eqnarray} \label{eq_12}
        	\sum_{r=1}^{z} \Gamma_r \left(\mathcal{X}_{n-r} \right)(t)  &=& \sum_{r=1}^{z}  z \displaystyle \int_{\frac{(r-1)h}{z}}^{\frac{rh}{z}}  \rho_{r} (xz - (r-1),t) \mathcal{X}_{n-r}(xz - (r-1)) \dif x \nonumber \\
        	&=&\int_{0}^{h} \left[z \sum_{r=1}^{z} \rho_{r} (sz - (r-1),t) \mathbbm{1}_{r}(s) \right] \left[\sum_{r=1}^{z} \mathcal{X}_{n-r}(sz - (r-1)) \mathbbm{1}_{r}(s) \right]\dif s, \nonumber
    	\end{eqnarray}
    	such that
    	\begin{equation} \label{eq_13a}
    	    \sum_{r=1}^{z} \Gamma_r \left(\mathcal{X}_{n-r} \right)(t) = \int_{0}^{h} \widetilde{\rho} (s,t) \widetilde{\mathcal{X}}_{n} (s) \dif s, \quad \widetilde{\rho} (s,t) = z \sum_{r=1}^{z} \rho_{r} (sz - (r-1),t) \mathbbm{1}_{r}(s),
    	\end{equation}
    	and $\widetilde{\mathcal{X}}_{n} (s) = \sum_{r=1}^{z} \mathcal{X}_{n-r}(sz - (r-1)) \mathbbm{1}_{r}(s)$, where we get, under \textbf{Assumption~\ref{as:2}}, the inequality $\int_{0}^{h} \int_{0}^{h}  \widetilde{\rho}^2 (s,t) \dif s \dif t  \leq z \sum_{r=1}^{z}  \int_{0}^{h}  \int_{0}^{h} \rho_{r}^2 (s,t) \dif s \dif t < \infty$, as long as $z < \infty$. Hence, we are under the FLMFR setting detailed in Section \ref{sec:flmfr}, under \textbf{Assumptions~\ref{as:1}--\ref{as:2}}, which ensure us the stationarity. Given a sample of a zero-mean ARH(z) process $\left\lbrace \mathcal{X}_i \right\rbrace_{i=0}^{n-1+z}$ valued in $\mathbb{H} = L^2 ([0,h])$, we then have, for each $i=1,\ldots,n$,
    	\begin{equation} \label{eq_14}
        	\widetilde{\mathcal{Y}}_i = \widetilde{\Gamma} \left(\widetilde{\mathcal{X}}_i \right) + \widetilde{\mathcal{E}}_i, \quad {\rm E} \left[\widetilde{\mathcal{E}}_i \mathrel{\big|} \widetilde{\mathcal{X}}_i \right] = 0, \quad \widetilde{\Gamma} (x) = {\rm E} \left[ \widetilde{\mathcal{Y}} \mathrel{\big|} \widetilde{\mathcal{X}} = x \right]\colon \mathbb{H} \mapsto \mathbb{H},
    	\end{equation}
    	where $\widetilde{\mathcal{X}}_i = \sum_{r=1}^{z} \mathcal{X}_{z-r+(i-1)} (sz - (r-1)) \mathbbm{1}_{r}(s)$, and $\widetilde{\mathcal{Y}}_i \coloneqq \mathcal{X}_{(i-1) + z}$ and $\widetilde{\mathcal{E}}_i \coloneqq \mathcal{E}_{(i-1) + z}$, for each $i=1,\ldots,n$, all of them centered. From~\eqref{eq_FLMFR_ARH}--\eqref{eq_14},
    	\begin{equation} \label{eq_13c}
    	    \widetilde{\Gamma}\left(\mathcal{X} \right)(\cdot) = \int_{0}^{h} \widetilde{\rho} (s,\cdot) \mathcal{X}(s) \dif s \quad \text{Hilbert-Schmidt operator}, \quad \widetilde{\rho} = \sum_{j=1}^{\infty} \sum_{k=1}^{\infty} \widetilde{\rho}_{jk} \Psi_j \otimes  \Psi_k,
    	\end{equation}
    	where  $\widetilde{\boldsymbol{\Gamma}}_{\widetilde{p},q} = \left(\widetilde{\rho}_{jk} \right)_{j=1,\ldots,\widetilde{p}}^{k=1,\ldots,q}$ is the $\widetilde{p} \times q$ matrix of the coefficients of $\widetilde{\rho}_{jk} = \langle \widetilde{\rho},  \Psi_j \otimes  \Psi_k \rangle_{\mathbb{H}^z}$, being $\widetilde{\rho}$ defined in~\eqref{eq_13a}. Now, $\left\lbrace \Psi_j \otimes \Psi_k \right\rbrace_{j,k=1}^{\widetilde{p},q}$ is constituted by the truncated versions of the same functional basis  $\left\lbrace \Psi_j  \right\rbrace_{j=1}^{\infty}$, although different cut-off levels $\left(\widetilde{p}, q \right)$ might be required. Note that, even though $\mathcal{X}_i$ and $\mathcal{E}_i$ are correlated, ${\rm E} \left[\widetilde{\mathcal{E}}_i \mathrel{\big|} \widetilde{\mathcal{X}}_i \right] = 0$, for each $i=1,\ldots,n$, since $\left(\mathcal{X}_{n-1}, \ldots, \mathcal{X}_{n-z}\right) $ and $\mathcal{E}_n$ are uncorrelated, for each $n \in \mathbb{Z}$. Since the previous $z$-lagged values are used, $\left\lbrace \mathcal{X}_i \right\rbrace_{i=0}^{n-1+z}$ are required for computing $\left\lbrace \widetilde{\mathcal{X}}_i \right\rbrace_{i=1}^{n}$.
    	
    	\medskip
    	\begin{remark}
    	Concerning to the case where $\widetilde{\Gamma}$ is furthermore compact, \textbf{Assumption~\ref{as:2}} could be modified, since $\widetilde{\Gamma}$ would admit a diagonal spectral decomposition in terms of an infinite sequence of real–valued AR(1) state equations, after projection the fully functional trajectories into the set of eigenfunctions of autocovariance operator (see, e.g., \citealp{SalmeronRuizMedina09} and \citealp{RuizMedinaSalmeron10}). In that case, we could extend in a direct way the asymptotic results by \cite{KoulStute99} for the real–valued case, such that considering ARH processes in terms of generalized processes (see \citealp{GelfandVilenkin64}), the weak–convergence of the projected empirical processes to a two–parameter generalized continuous Gaussian process may be achieved. This process would admit in distribution sense a representation, in terms of two–parameter generalized Brownian motion.
    	\end{remark}

	\subsection{A GoF test for ARH(z) modelss}
	\label{sec:gof_ARHp}

        As exposed, the guiding thread of this paper is to verify whether the relation between frv's $\left\lbrace \widetilde{\mathcal{Y}}_i= \mathcal{X}_{i-1 + z} \right\rbrace_{i=1}^{n}$ and their $z$-lagged values $\left\lbrace \left(\mathcal{X}_{z - 1 + (i-1)}, \mathcal{X}_{z - 2 + (i-1)}, \ldots, \mathcal{X}_{ (i-1)} \right) \right\rbrace_{i=1}^{n}$ can be linearly related. For this purpose, a GoF test for the ARH($z$) model in \eqref{eq_8} is now formulated, based on its characterization as a FLMFR. Particularly, for a given $z \geq 1$, we want to test the composite null hypothesis $\mathcal{H}_0$ (against an unspecified alternative)
    	\begin{equation} \label{eq_18b}
        	\mathcal{H}_0\colon \mathcal{X}_n \text{ and } \left(\mathcal{X}_{n-1}, \ldots, \mathcal{X}_{n-z} \right) \text{ are linearly related}, \quad n \in \mathbb{Z}.
    	\end{equation}
    	Note the reader that, from   Proposition \ref{prop:1}, the null hypothesis $\mathcal{H}_0$ in \eqref{eq_18b} is equivalent to
    	\begin{equation} \label{eq_19b}
    	\mathcal{H}_0\colon \overline{\mathcal{X}}_n = \overline{\Gamma} \left(\overline{\mathcal{X}}_{n-1} \right) + \overline{\mathcal{E}}_n, \quad   \overline{\Gamma}\left(\overline{\mathcal{X}} \right) (t) \coloneqq \overline{\Gamma}_{\overline{\rho}} \left(\overline{\mathcal{X}} \right) (t) = \left\langle  \overline{\rho}(\cdot, t), \overline{\mathcal{X}}(\cdot) \right\rangle_{\mathbb{H}^z}, \quad \overline{\mathcal{X}} \in \mathbb{H}^z, 
    	\end{equation}
    	with $n \in \mathbb{Z}$, for some unknown squared-integrable  $\overline{\rho} \in \mathbb{H}^z \otimes_{\mathbb{H}^z} \mathbb{H}^z$, where $\mathbb{H}^z$ is a separable Hilbert space, as proved in Lemma 1 in the Supplementary Material. Furthermore, from  Section \ref{sec:arh_as_flmfr},   $\mathcal{H}_0$ in \eqref{eq_18b}--\eqref{eq_19b} is also equivalent to
    	\begin{equation} \label{eq_20} 
    	\mathcal{H}_0\colon \widetilde{\Gamma} \in \mathcal{L} = \left\lbrace \langle \langle \cdot, \widetilde{\rho} \rangle \rangle\colon \widetilde{\rho} \in \mathbb{H} \otimes \mathbb{H} \right\rbrace, \quad \widetilde{\mathcal{Y}} = \widetilde{\Gamma} \left(\widetilde{\mathcal{X}} \right) + \widetilde{\mathcal{E}}, \quad   \widetilde{\Gamma}_{\widetilde{\rho}} \left(\widetilde{\mathcal{X}} \right) (t) = \int_{0}^{h} \widetilde{\rho}(s, t) \widetilde{\mathcal{X}}(s) \dif s,
    	\end{equation}
    	where $\widetilde{\mathcal{X}}_i = \sum_{r=1}^{z} \mathcal{X}_{z-r+(i-1)} (sz - (r-1)) \mathbbm{1}_{r}(s)$, and $\widetilde{\mathcal{Y}}_i:= \mathcal{X}_{i-1 + z}$ and $\widetilde{\mathcal{E}}_i:= \mathcal{E}_{i-1 + z}$ centered  frv's, for each $i=1,\ldots,n$, being $\widetilde{\Gamma}$ a Hilbert-Schmidt integral operator. In what follows, we will consider the characterization of null hypothesis $\mathcal{H}_0$ in~\eqref{eq_20} from a FLMFR perspective, based on developments in \eqref{eq_FLMFR_ARH}--\eqref{eq_13c}.

    	From \citeauthor{GarciaPortuguesetal20} (\citeyear[Lemmas 1-2]{GarciaPortuguesetal20}), the null hypothesis in~\eqref{eq_20} can be also characterized in terms of the projections of the functional paths as follows
    	$$\mathcal{H}_0\colon~ {\rm E} \left[ \langle \widetilde{\mathcal{Y}} - \langle \langle \widetilde{\mathcal{X}}, \widetilde{\rho} \rangle\rangle, \gamma_{\widetilde{\mathcal{Y}}} \rangle_{\mathbb{H}} \mathbbm{1}_{\left\lbrace \langle \widetilde{\mathcal{X}}, \gamma_{\widetilde{\mathcal{X}}} \rangle_{\mathbb{H}} \leq u \right\rbrace} \right] = 0, \quad \text{for almost every } u \in \mathbb{R},$$
    	and for each direction $\gamma_{\widetilde{\mathcal{X}}}, \gamma_{\widetilde{\mathcal{Y}}} \in \mathbb{S}_{\mathbb{H}}$, 
    	where $\mathbb{S}_{\mathbb{H}} = \left\lbrace \mathcal{X} \in \mathbb{H}:~\left\| \mathcal{X} \right\|_{\mathbb{H}} = 1 \right\rbrace$  is the functional analogue of the Euclidean sphere. Extending the ideas of \cite{Escanciano06}, deviations from $\mathcal{H}_0$ can be detected by computing a residual marked empirical process
    	\begin{equation}
    	R_n \left(u, \gamma_{\widetilde{\mathcal{X}}}, \gamma_{\widetilde{\mathcal{Y}}} \right) = \frac{1}{\sqrt{n}} \sum_{i=1}^{n} \langle \widetilde{\mathcal{E}}_i,  \gamma_{\widetilde{\mathcal{Y}}}  \rangle_{\mathbb{H}} \mathbbm{1}_{\left\lbrace \langle \widetilde{\mathcal{X}}, \gamma_{\widetilde{\mathcal{X}}} \rangle_{\mathbb{H}} \leq u \right\rbrace}, \quad u \in \mathbb{R}, \quad \gamma_{\widetilde{\mathcal{X}}}, \gamma_{\widetilde{\mathcal{Y}}}\in \mathbb{S}_{\mathbb{H}}, \nonumber
    	\end{equation} 
    	where marks are given by the projected errors  $\left\lbrace \langle \widetilde{\mathcal{E}}_i, \gamma_{\widetilde{\mathcal{Y}}} \rangle_{\mathbb{H}} = \langle \widetilde{\mathcal{Y}}_i - \langle \langle \widetilde{\mathcal{X}}_i, \hat{\rho} \rangle\rangle , \gamma_{\widetilde{\mathcal{Y}}} \rangle_{\mathbb{H}} \right\rbrace_{i=1}^{n}$ and jumps are determined by the projected regressors $\left\lbrace \langle \widetilde{\mathcal{X}}_i, \gamma_{\widetilde{\mathcal{X}}} \rangle_{\mathbb{H}} \right\rbrace_{i=1}^{n}$, where $\hat{\rho}$ denotes some estimator of $\widetilde{\rho}$ (see equations~\eqref{eq_FLMFR_ARH}--\eqref{eq_13c}). To measure how close the empirical process  is to zero,  a  projected Cramér–von Mises (PCvM) statistic, on the norm on  $\mathbb{S}_{\mathbb{H}} \times \mathbb{S}_{\mathbb{H}} \times \mathbb{R}$, will be adopted as
    	$$\text{PCvM}_n = \int_{\mathbb{S}_{\mathbb{H}} \times \mathbb{S}_{\mathbb{H}} \times \mathbb{R}} \left[ R_n \left(u, \gamma_{\widetilde{\mathcal{X}}}, \gamma_{\widetilde{\mathcal{Y}}} \right) \right]^2 F_{n, \gamma_{\widetilde{\mathcal{X}}}} (\dif u) \omega_{\widetilde{\mathcal{X}}} (\dif \gamma_{\widetilde{\mathcal{X}}} ) \omega_{\widetilde{\mathcal{Y}}} (\dif \gamma_{\widetilde{\mathcal{Y}}} ),$$
    	where $F_{n, \gamma_{\widetilde{\mathcal{X}}}}$ is the empirical cumulative distribution function (ecdf) of projected regressors, and $\omega_{\widetilde{\mathcal{X}}}$ and $\omega_{\widetilde{\mathcal{Y}}}$ are suitable measures on $\mathbb{S}_{\mathbb{H}}$, respectively.  The infinite dimension of  functional spheres makes this  functional  hard to handle. For that reason, the projected Cramér-von Mises statistic will be expressed in terms of finite-dimensional directions $\gamma_{\widetilde{\mathcal{X}}}^{(\widetilde{p})}$ and $\gamma_{\widetilde{\mathcal{Y}}}^{(q)}$, as well as $\widetilde{\mathcal{E}}_{i}^{(q)}$ and  $\widetilde{\mathcal{X}}^{(\widetilde{p})}$, just adopting a ($\widetilde{p}$, q)-truncated expansions. After some simple algebra developments, an easily computable statistic was obtained as
    	\begin{equation} \label{eq_18} 
    	\text{PCvM}_{n,\widetilde{p},q} = \frac{1}{n^2} \frac{2 \pi^{\widetilde{p}/2 + q/2 - 1}}{q \Gamma(\widetilde{p}/2)  \Gamma(q/2)} \text{Tr} \left[\widetilde{\textbf{E}}_{q}^{T} \textbf{A}_{\bullet} \widetilde{\textbf{E}}_{q} \right], \quad  \left(\widetilde{\textbf{E}}_q \right)_{i = 1,\ldots,n}^{k=1,\ldots,q} = \hat{e}_{i,k}  = \langle \widetilde{\mathcal{E}}_i, \Psi_k \rangle_{\mathbb{H}},
    	\end{equation}
    	being $\text{Tr} \left[\cdot \right]$ the trace operator,  $\Gamma \left( \cdot \right)$ the Gamma function and $\widetilde{\textbf{E}}_q$ the error coefficients, that is, the projected estimators of $\widetilde{\mathcal{E}}_i:= \mathcal{E}_{i-1 + z}$, for each $i=1,\ldots,n$.  The matrix $\textbf{A}_{\bullet} = \left(\textbf{A}_{\bullet} \right)_{ij} = \left( \sum_{r=1}^{n} A_{ijr} \right)_{ij} $ is obtained as surface areas of particular spherical regions, which explicit expression can be given as (see \citealp{GarciaPortuguesetal14} and \citealp{GarciaPortuguesetal20})
    	\begin{equation}
    	    A_{ijr} = A_{ijr}^{\measuredangle} \frac{\pi^{\widetilde{p}/2 - 1}}{\Gamma(\widetilde{p}/2)}, \quad A_{ijr}^{\measuredangle}:=
    	        \begin{cases}
        	        2 \pi, \quad \boldsymbol{x}_{i,\widetilde{p}} = \boldsymbol{x}_{j,\widetilde{p}} = \boldsymbol{x}_{r,\widetilde{p}} \\
        	        \pi, \quad \boldsymbol{x}_{i,\widetilde{p}} \neq \boldsymbol{x}_{j,\widetilde{p}} \text{ , and } \boldsymbol{x}_{i,\widetilde{p}} = \boldsymbol{x}_{r,\widetilde{p}} \text{ or } \boldsymbol{x}_{j,p} = \boldsymbol{x}_{i,\widetilde{p}}\\
        	        \pi - \cos^{-1} \left( \frac{\left( \boldsymbol{x}_{i,\widetilde{p}} - \boldsymbol{x}_{r,\widetilde{p}} \right)^{T} \left(\boldsymbol{x}_{j,\widetilde{p}} - \boldsymbol{x}_{r, \widetilde{p}}\right) }{ \left\| \boldsymbol{x}_{i, \widetilde{p}} - \boldsymbol{x}_{r, \widetilde{p}}\right\| \left\| \boldsymbol{x}_{j, \widetilde{p}} - \boldsymbol{x}_{r, \widetilde{p}}\right\| }\right),
                \end{cases}
    	\end{equation}
    	where $\boldsymbol{x}_{i,\widetilde{p}} = \left(\langle \mathcal{X}_i, \Psi_1 \rangle, \ldots, \langle \mathcal{X}_i, \Psi_{\widetilde{p}} \rangle \right)^{T}$. Geometrical arguments $\left(A_{ijr}\right)_{i,j,r = 1,\ldots,n}$ can be directly computed from the $\textsf{R}$ package \textsf{goffda} \citep{GarciaPortuguesAlvarezLiebana19}, and they depend exclusively on the covariates, and then, they only needs to be computed once in the testing procedure. The GoF test for ARH(z) models proposed is then detailed within the next algorithm.
    	
	\begin{algorithm}[GoF test in practice] \label{al:ARHz}
		Let $\left\lbrace \mathcal{X}_i \right\rbrace_{i=0}^{n-1+z}$ be a sample of a centered ARH(z) model,  under \textbf{Assumptions~\ref{as:1}--\ref{as:2}} and the setting above detailed. We proceed as follows:
		\begin{enumerate}
			\item Construct frv's $\widetilde{\mathcal{X}}_i (s)\coloneqq \sum_{r=1}^{z} \mathcal{X}_{z-r+(i-1)} (sz - (r-1)) \mathbbm{1}_{r}(s)$,  $\widetilde{\mathcal{Y}}_i (t)\coloneqq \mathcal{X}_{(i-1) + z}(t)$ and $\widetilde{\mathcal{E}}_i (t)\coloneqq \mathcal{E}_{(i-1) + z}(t)$, all of them centered, with $\widetilde{\mathcal{Y}}_i = \widetilde{\Gamma} \left(\widetilde{\mathcal{X}}_i \right) + \widetilde{\mathcal{E}}_i$, for any $i=1,\ldots,n$, and $s,t \in [0,h]$, constituting a particular case of FLMFR.
			\item Compute the FPC of $\left\lbrace \widetilde{\mathcal{X}}_i  \right\rbrace_{i=1}^{n}$ and $\left\lbrace \widetilde{\mathcal{Y}}_i  \right\rbrace_{i=1}^{n}$, and choose initial cut-off levels $(p,q)$ as the minimum number of FPC required for capturing a proportion of $\text{EV}$ (say $\text{EV}_p$ = $\text{EV}_q$ = 0.99). After that, we obtain the $p$- and $q$-truncated FPC scores $\widetilde{\mathbf{X}}_p$  and $\widetilde{\mathbf{Y}}_q$ of $\left\lbrace \widetilde{\mathcal{X}}_i  \right\rbrace_{i=1}^{n}$ and $\left\lbrace \widetilde{\mathcal{Y}}_i  \right\rbrace_{i=1}^{n}$, respectively.
			\item Compute the FPCR-L1S estimator $\widehat{\widetilde{\mathbf{B}}}_{\widetilde{p},q}^{(\lambda), C}$ of $\widetilde{\mathbf{B}}_{\widetilde{p},q}$, where $\widetilde{\mathbf{B}}_{\widetilde{p},q} = \left(\widetilde{\beta}_{jk} \right)_{j=1,\ldots,\widetilde{p}}^{j=1,\ldots,q}$ is the $\widetilde{p} \times q$ matrix of the coefficients of $\widetilde{\beta}_{jk} = \langle \widetilde{\rho},  \Psi_j \otimes  \Psi_k \rangle_{\mathbb{H}^z}$, and the associated coefficients of residuals $\widetilde{\mathbf{e}}_{i,q} = \widetilde{\mathbf{Y}}_{i,q} - \widetilde{\mathbf{X}}_{i,\widetilde{p}}\widehat{\widetilde{\mathbf{B}}}_{\widetilde{p},q}^{(\lambda), C}$, for each $i=1,\ldots,n$. Note that $\widetilde{p}$ out of $p$ FPC coefficients are automatically selected. We compute the statistic $\text{PCvM}_{n,\widetilde{p},q}$ in \eqref{eq_18}.
			\item Implement a golden-section bootstrap, by simulating independent zero-mean and unit-variance random variables, setting  bootstrapped errors as $\mathbf{\widetilde{e}}_{i,q}^{\ast b}$ and $\widetilde{\mathbf{Y}}_{i,q}^{\ast b}\coloneqq \widetilde{\mathbf{X}}_{i,\widetilde{p}}\widehat{\widetilde{\mathbf{B}}}_{\widetilde{p},q}^{(\lambda), C} + \mathbf{\widetilde{e}}_{i,q}^{\ast b}$, for each $i=1,\ldots,n$ and $b=1,\ldots,B$, being $B$ the number of replicates.
			\item After centering them,  from the bootstrap sample, recompute the FPCR-L1S estimator in \eqref{eq_4}, obtaining  the bootstrapped residuals and statistic $\text{PCvM}_{n,\widetilde{p},q}^{\ast b}$ from \eqref{eq_18}.
			\item Estimate the p-value  as $\#\left\lbrace \text{PCvM}_{n,\widetilde{p},q} \leq  \text{PCvM}_{n,\widetilde{p},q}^{\ast b}\right\rbrace / B$.
		\end{enumerate}
		
	\end{algorithm}

	\subsection{Numerical results: a comparative study}
	\label{sec:sim_ARHp}
	
    The performance of the test, regarding power and size, is now compared to the multi-stage testing procedure in \cite{KokoszkaReimherr12}, abbreviated as \textbf{KR}, to determine the order of an ARH process. Note that their approach was developed just under ARH alternatives, in contrast with our unspecified alternatives. The following common settings will be used through scenarios described in Table~\ref{tab: scenariosARH}: functional trajectories $\left\lbrace \mathcal{X}_i \right\rbrace_{i=1}^{n}$ and errors $\left\lbrace \mathcal{E}_i \right\rbrace_{i=1}^{n}$ (standard Brownian bridges), as shown in Figure~\ref{fig:1}, are valued in 101 equispaced points in $[0,1]$, with $\text{M} = 1\,000$ Monte Carlo replicates, $\text{B} = 1\,500$ bootstrap resamples.  The initial values were simulated as standard Brownian bridges and a burn-in period of 200 functional observations was used. Our test was run following Algorithm~\ref{al:ARHz}, using initial cut-off levels $(p, q)$ for ensuring us an explained variance of $\text{EV}_p = \text{EV}_q = 0.995$. Subsequently, the FPCR-L1S estimator in \eqref{eq_4} was implemented by using the so-called one standard error rule $\lambda_{1SE}$, such that a L1-based variable selection is achieved. Note that, as discussed in Section 2 in the Supplementary Material, the variable selection is addressed focusing on the testing procedure, not on the estimation of the autocorrelation operator.
	\begin{table}[H]
	\centering
	\small
	\renewcommand{\arraystretch}{0.6}
	\caption{Summary of null and alternative hypotheses.}
	\begin{tabular}{cll}
		\toprule
	    Notation & \multicolumn{1}{c}{Model} & \multicolumn{1}{c}{Parameters}  \\
		\midrule
		\textbf{ARH(0)} & $\mathcal{X}_n (t) = \mathcal{E}_n (t)$ & None  \\
		\hline
		\multirow{2}[2]{*}{\textbf{ARH(1)}} & $\mathcal{X}_n (t) = \Gamma_1 (\mathcal{X}_{n-1}) (t) + \mathcal{E}_n (t)$ & \multirow{2}{*}{$\left\| \rho_1 \right\|_{L^2 \left([0, 1]^2 \right)} = 0.7$}  \\
		& $ \rho_1 (s, t) = c_{1} (2 - (2s-1)^2 - (2t-1)^2) $ & \\
		\hline
		\multirow{2}[2]{*}{\textbf{ARH(2)}} & $\mathcal{X}_n (t) = \Gamma_1 (\mathcal{X}_{n-1}) (t) + \Gamma_2 (\mathcal{X}_{n-2}) (t) + \mathcal{E}_n (t)$ & $\left\| \rho_1 \right\|_{L^2 \left([0, 1]^2 \right)} = 0.5$  \\
		& $ \rho_i (s, t) = c_{2, i} e^{-(t^2+s^2)/2} $ & $\left\| \rho_2 \right\|_{L^2 \left([0, 1]^2 \right)} = 0.3$  \\
		\hline
		Non linear, & $\mathcal{X}_n (t) =  \Gamma_1 \left( \mathcal{X}^2_{n-1} \right) (t)  + \mathcal{E}_n (t)$ & \multirow{2}[2]{*}{$\left\| \rho_1 \right\|_{L^2 \left([0, 1]^2 \right)} = 0.5$}  \\
		quadratic (\textbf{NLQ}) & $ \rho_1 (s, t) = c_{2, 1} e^{-(t^2+s^2)/2} $ &  \\
		\hline
		Non linear,  & $\mathcal{X}_n (t) = \Gamma_1 \left( \left| \mathcal{X}_{n-1} \right|^{0.5}  \right) (t) + \mathcal{E}_n (t)$ & \multirow{2}[2]{*}{$\left\| \rho_1 \right\|_{L^2 \left([0, 1]^2 \right)} = 0.5$}  \\
		square root (\textbf{NLS}) & $ \rho_1 (s,t) = c_{2, 1}  e^{-(t^2+s^2)/2} $ &  \\
		\bottomrule
	\end{tabular}%
	\label{tab: scenariosARH}
	\end{table}
	As summarized in Table~\ref{tab: scenariosARH}, we generated ARH($z$) models, with  $z \in \lbrace 0, 1, 2 \rbrace$, and two nonlinear models, where $\Gamma \left(\mathcal{X} \right) (\cdot) = \int_{0}^{1} \rho(s, \cdot) \mathcal{X}(s) \dif s$ and $\rho$  are parabolic or Gaussian kernels, commonly used in the literature (see \citealp{Gabrysetal10} and \citealp{Hormannetal10}). In Table~\ref{tab: scenariosARH},  $\mathcal{X}^2$ and $\left| \mathcal{X} \right|^{0.5}$ denote the pointwise exponentiation and square root operations, respectively, to the discretized trajectories. Similar conditions on $\left\| \rho \right\|_{L^2 \left([0, 1]^2 \right)}$ proposed in \cite{KokoszkaReimherr12} are imposed, such that \textbf{Assumptions~\ref{as:1}--\ref{as:2}} are held, and $\left(c_{1}, c_{2, 1}, c_{2, 2} \right) =\left( 0.500568, 0.669502, 0.401701 \right)$ are set. Table~\ref{tab:powerARH} shows the empirical rejection rates of testing $\mathcal{H}_0$ whether $\mathcal{X}_n$ is an ARH process of order $z$, with $z \in \left\lbrace 0, 1\right\rbrace$, comparing the performance of our proposed \textbf{CvM} test and the \textbf{KR} test, for scenarios in which data is generated following an ARH($z$) model, with $z \in \left\lbrace 0, 1, 2\right\rbrace$, and two nonlinear models. As remarked, the former test  does not specify an alternative hypothesis, while the latter assumes a specific alternative (testing an ARH($z$) against an ARH($z+1$) model). Nominal level $\alpha=0.05$ and sample sizes $n \in \lbrace 150, 250, 350, 500 \rbrace$, are considered.
	\begin{table}[t]
		\centering
		{\small
		\renewcommand{\arraystretch}{0.6}
	\caption{\label{tab:powerARH}Empirical size and power of \textbf{CvM} and  \textbf{KR} tests for testing the ARH process order. 
	Under $\mathcal{H}_0$, boldfaced rejection rates lie outside a 95\%-confidence interval for $\alpha = 0.05$.}
		\begin{tabular}{ccccccccccc}
		    \toprule
			\multicolumn{1}{l}{} & \multicolumn{1}{l}{} & \multicolumn{4}{c}{\textbf{CvM} [unspecified $\mathcal{H}_1$]}   && \multicolumn{4}{c}{\textbf{KR}} \\
			\cmidrule{3-6} \cmidrule{8-11}   
			Data & $\mathcal{H}_0$ & 150 & 250 & 350 & 500 & $\mathcal{H}_1$ & 150 & 250 & 350 & 500 \\
			\midrule
			\multirow{2}[2]{*}{\textbf{ARH(0)}}
			 & ARH(0) & 0.042 & 0.042 & 0.047 & 0.041 & ARH(1) & 0.045 & 0.056 & 0.051 & 0.054 \\
			 & ARH(1) & 0.045 & 0.042 & 0.047 & 0.040 & ARH(2) & 0.056 & 0.055 & 0.061 & 0.050 \\
			\hline
			\multirow{2}[2]{*}{\textbf{ARH(1)}} 
			& ARH(0) & 1.000 & 1.000 & 1.000 & 1.000 & ARH(1) & 1.000 & 1.000 & 1.000 & 1.000 \\
			& ARH(1) & \textbf{0.030} & 0.047 & 0.058 & 0.041 & ARH(2) & 0.056 & 0.053 & 0.063 & 0.042 \\
			\hline
			\multirow{2}[2]{*}{\textbf{ARH(2)}} 
			 & ARH(0) & 1.000 & 1.000 & 1.000 & 1.000 & ARH(1) & 1.000 & 1.000 & 1.000 & 1.000 \\
			 & ARH(1) & 0.367 & 0.596 & 0.632 & 0.651 & ARH(2) & 0.954 & 1.000 & 1.000 & 1.000 \\
			\hline
			\multirow{2}[2]{*}{\textbf{NLQ}}
			 & ARH(0) & 0.541 & 0.818 & 0.949 & 0.998 & ARH(1) & 0.406 & 0.652 & 0.842 & 0.972 \\
			 & ARH(1) & 0.535 & 0.808 & 0.946 & 0.984 & ARH(2) & 0.059 & 0.047 & 0.058 & 0.048 \\
			\hline
			\multirow{2}[2]{*}{\textbf{NLS}} 
			 & ARH(0) & 0.203 & 0.358 & 0.580 & 0.795 & ARH(1) & 0.164 & 0.235 & 0.311 & 0.426 \\
			 & ARH(1) & 0.198 & 0.355 & 0.597 & 0.797 & ARH(2) & 0.056 & 0.045 & 0.069 & 0.053 \\
			\bottomrule
		\end{tabular}%
		}
	\end{table}%
	%
	According to scenarios displayed in Table~\ref{tab:powerARH}, regarding the size under null hypotheses (first, second and forth row), \textbf{CvM} and \textbf{KR} tests seem to be well calibrated, though both show a slight over-rejection in the ARH(1) scenario when indeed testing ARH(1) as $\mathcal{H}_0$. As observed, size of tests are also close to the nominal level $\alpha = 0.05$, even when small sample size are considered. Concerning powers under ARH alternatives (third, fifth and sixth rows), the \textbf{KR} test is more powerful, as expected since the \textbf{KR} test was specifically designed for these scenarios, rejecting ARH($z$) in favor of ARH($z+1$). However, since \textbf{KR} test assumes an ARH structure  even under alternatives, it fails against nonlinear alternatives (eighth, ninth and tenth rows), showing a poor power, in contrast with the high empirical power exhibits by our \textbf{CvM test} in most of scenarios.  As opposed to the \textbf{KR} test, our proposal shows an increasing power inasmuch as increasing the sample size.
	\section{Stochastic differential equations: specification test}
	\label{sec:sde}
	
	The methodology introduced in Section~\ref{sec:GoF} can be further extended to diffusion processes, where the common framework is one dimensional, as usually one path of the process is considered.
	However, in finance, and motivated by the increasingly availability of high-frequency data, multivariate GoF tests could be unsuitable for such data, and thus, functional data setup has been exploited to take advantage of the shapes of curves \citep{Mulleretal11}. As a sideways contribution, we now propose a new specification test for diffusion models from a high-dimensional data perspective,
	namely for OU processes. 

	\subsection{Diffusion models}
	    As referred, stochastic diffusion models, as solutions to SDEs, have been increasingly used the last decades. 
	    From now on, given a filtered space $\big(\Omega, \left\lbrace \mathcal{A}_t \right\rbrace_{t \in \left[0,T \right]}, \mathbb{P} \big)$, associated to a probability space, we consider the parametric time-homogeneous one dimensional SDE
    	\begin{equation}\label{homogeneous_SDE}
    	\dif \xi_t = m(\xi_t,  \boldsymbol{\theta}) \dif t + \sigma (\xi_t,  \boldsymbol{\theta}) \dif W_t, \quad t \in \mathbb{R}^{+}, \quad \left\lbrace W_t \right\rbrace_{t \in \mathbb{R}^+}~\text{a standard Wiener process}, 
    	\end{equation} 
    	where $m\colon\mathbb{R}\times \mathbb{R}^d \to \mathbb{R}$ and $\sigma\colon\mathbb{R}\times \mathbb{R}^d \to \mathbb{R}^{+}$ are the drift and volatility functions, respectively, with $\boldsymbol{\theta} \in \Theta \subset \mathbb{R}^d$. The diffusion model in~\eqref{homogeneous_SDE} is commonly represented in its integral version, where $\xi_t$ is $\mathcal{A}_t-$measurable, for each $t \in [0,T]$, as
    	\begin{equation} \label{integral_SDE}
    	\xi_t = \xi_0 + \int_{0}^{T} m(\xi_s, \boldsymbol{\theta}) \dif s + \int_{0}^{T} \sigma (\xi_s, \boldsymbol{\theta}) \dif W_s, \qquad \xi_0~\text{ initial condition at}~t_0 = 0,
    	\end{equation}
    	%
    A strong solution of~\eqref{integral_SDE} is guaranteed under the boundedness, and Lipschitz and linear growth conditions (see, e.g., \citealp{KaratzasShreve98}) on $m(\cdot)$ and $\sigma(\cdot)$:

    	\medskip
    	
    	\begin{assumption}[Global Lipschitz] \label{as:41}
    		For all $x, y \in \mathbb{R}$, there exist $C_1 \in \left(0, \infty \right)$, not depending on parameters $\boldsymbol{\theta}$, such that $\left|m(x, \boldsymbol{\theta})-m(y, \boldsymbol{\theta})\right| + \left|\sigma(x, \boldsymbol{\theta})-\sigma(y, \boldsymbol{\theta})\right| \leq C_1 \left|x-y\right|$.
    	\end{assumption}
    	\begin{assumption}[Linear growth]\label{as:42}
    		For all $x \in \mathbb{R}$, there exist $C_2 \in \left(0, \infty \right)$, not depending on the set of parameters $\boldsymbol{\theta}$, such that $\left| m(x, \boldsymbol{\theta})\right| + \left|\sigma (x, \boldsymbol{\theta})\right| \leq C_2 \left(1 + \left| x \right| \right)$.
    	\end{assumption}
    	
    	\medskip
    	
    	SDEs  are usually discretely observed on a time interval $[0, T]$, discretized at equally spaced time points $\left\lbrace t_0, t_1, \dots, t_N\right\rbrace$. In keeping with the Euler-Maruyama  discretization scheme, we approximate SDE in~\eqref{homogeneous_SDE} as $\xi_{t_{i+1}} - \xi_{t_i} \approx m \left(\xi_{t_i}, \boldsymbol{\theta} \right) \Delta + \sigma \left(\xi_{t_i}, \boldsymbol{\theta} \right) \sqrt{\Delta} \left(W_{t_{i+1}} - W_{t_i}\right)$, for each $i=0,1,\dots,N-1$,
    	where $\Delta = T/N$ is the length of the sampling intervals, and	$\left\lbrace \left(W_{t_{i+1}} - W_{t_i}\right) \right\rbrace_{i=0}^{N-1}$ are iid zero-mean Gaussian random variables. 
    	Among the different specifications of~\eqref{homogeneous_SDE}, a common family of parametric models have been extensively used during the last decades, motivated by capturing the dynamics of the short-term interest rates (see, e.g., \citealt{Vasicek77}), which can be nested in the CKLS model \citep{chan1992empirical},
    \begin{equation} \label{eq:ckls}
    \dif \xi_t = \kappa(\mu - \xi_t) \dif t + \sigma \xi_t^\gamma \dif W_t, \qquad t \in \mathbb{R}^{+}, \qquad \sigma > 0~\text{(volatility around the mean)},
    \end{equation}
    where $\mu$ is the long term mean and $\kappa > 0$ is the speed of adjustment to $\mu$ (rate of mean reversion). A wide class of interest rate models can be obtained from~\eqref{eq:ckls}, imposing restrictions to $\left(\mu, \kappa, \sigma, \gamma \right)$, mainly epitomized by  well-known the OU processes, with $\gamma = 0$.
	\subsection{Ornstein-Uhlenbeck: a particular ARH(1) process}
	\label{sec:ou}
    As referred, the relevance of OU processes is not limited to finance, since they have a long history in physics 
    This process, which constitutes a CKLS model in~\eqref{eq:ckls} with $\gamma = 0$, and formalized as $\dif \xi_t = \kappa(\mu - \xi_t) \dif t + \sigma \dif W_t$, with $t \in \mathbb{R}^{+}$, and $\kappa,~\sigma > 0$, 
    	is often employed under its integral representation (via Fourier method of separation of variables), with $\kappa,~\sigma > 0$:
    	\begin{equation}\label{integral_OU}
        	\xi_{t} = \xi_0  e^{-\kappa t}  + \mu \left( 1 -  e^{-\kappa t} \right) + \sigma \int_{0}^{t} e^{-\kappa (t-s)} \dif W_s, \quad t \in \mathbb{R}^{+},\; \xi_0~\text{initial condition at}~t_0 = 0.
	    \end{equation}
	    With the same philosophy of the splitted representation adopted in Section~\ref{sec:GoF} (see Figure~\ref{fig:1}), and according to the proposal by \cite{AlvarezLiebanaetal16}, the OU process in~\eqref{integral_OU} can be characterized as an ARH(1) process. Let $\mathbb{H}$ be a separable Hilbert space given by $\mathbb{H} = L^2 \left([0,h], \mathcal{B}_{[0,h]}, \lambda + \delta_{(h)}\right)$, where $\mathcal{B}_{[0,h]}$ is the  $\sigma$-algebra generated by subintervals $[0, h]$, $\lambda$ denotes the Lebesgue measure and $\delta_{(h)} (s) =\delta (s-h)$ is the Dirac measure at $h$. The associated norm is defined as $\left\| \mathcal{X} \right\|_{\mathbb{H}} = \sqrt{\int_{0}^{h} \mathcal{X}^2 (s) \dif \left(\lambda + \delta_{(h)} \right)} = \sqrt{\int_{0}^{h} \mathcal{X}^2 (s) \dif s + \mathcal{X}^2 (h)}$. Remark that $\left\| \cdot \right\|_{\mathbb{H}}$ directly establishes equivalent classes of functions such that $\mathcal{X} \sim_{\lambda + \delta_{(h)}} \mathcal{Y}$ if and only if $\lambda \left( \left\lbrace s\colon \mathcal{X} (s) \neq \mathcal{Y}(s) \right\rbrace \right) = 0$ and $\mathcal{X} (h) = \mathcal{Y}(h)$. In this setting, the OU process in~\eqref{integral_OU} can be splitted as follows, for each $n \in \mathbb{Z}$:
    	\begin{equation*}
        	\mathcal{X}_n (t) = \xi_{nh+t} = \xi_0 e^{-\kappa \left(nh + t \right)} + \mu \left(1 - e^{-\kappa \left(nh + t \right)} \right) + \sigma \int_{0}^{nh+t} e^{-\kappa (nh+t-s)} \dif W_s, \qquad t \in [0,h],
    	\end{equation*}
    	and then, $	e^{-\kappa t} \mathcal{X}_{n-1} (h) = 	e^{-\kappa t} \xi_{nh}  = \xi_0 e^{-\kappa \left(nh + t \right)} + \mu \left( e^{-\kappa t}  - e^{-\kappa \left(nh + t \right)} \right) + \sigma \int_{0}^{nh} e^{-\kappa (nh +t -s)} \dif W_s$. The set of paths $\left\lbrace \mathcal{X}_n (t) \right\rbrace_{n \in \mathbb{Z}} = \left\lbrace \xi_{nh + t}\colon t \in \left[0, h \right] \right\rbrace_{n \in \mathbb{Z}}$ can be expressed as follows:
    	\begin{equation*}
    	e^{-\kappa t} \left( \mathcal{X}_{n-1} (h) - \mu \right) = \left( \mathcal{X}_n(t) - \mu \right) - \sigma \int_{nh}^{nh + t} e^{-\kappa (nh +t -s)} \dif W_s.
    	\end{equation*}
    	
    	From now on, $\left\lbrace \xi_t \right\rbrace_{t \in \mathbb{R}^{+}}$ will be centered respect to its long term mean ($\mu = 0$), and, from \cite{AlvarezLiebanaetal16}, the OU process in~\eqref{integral_OU} can be characterized as a zero-mean stationary ARH(1) process $\left\lbrace \mathcal{X}_n (t) \coloneqq \xi_{nh+t},~t \in \left[0, h \right] \right\rbrace_{n \in \mathbb{Z}}$ (see Figure~\ref{fig:1}), given by
    	\begin{equation}\label{arh_ou}
    	\mathcal{X}_n(t) = e^{-\kappa t} \mathcal{X}_{n-1} (h) + \sigma \int_{nh}^{nh + t} e^{-\kappa (nh +t -s)} \dif W_s = \Gamma_{\kappa} \left( \mathcal{X}_{n-1} \right) (t) + \mathcal{E}_n \left(t \right), \qquad n \in \mathbb{Z},
    	\end{equation}
    	where $\left\lbrace \mathcal{E}_n (t) \coloneqq \sigma \int_{nh}^{nh + t} e^{-\kappa (nh +t -s)} \dif W_s \right\rbrace_{n \in \mathbb{Z}}$ constitutes a $\mathbb{H}$-valued strong white noise and  $\Gamma_{\kappa}$ is a bounded linear operator, for each $\kappa > 0$ \citep{AlvarezLiebanaetal16}. Note the reader that $\left\| \Gamma_{\kappa} \left( \mathcal{X} \right) \right\|_{\mathbb{H}}^2 = \int_{0}^{h} \left| \Gamma_{\kappa} \left( \mathcal{X} \right) (t) \right|^2 \dif t +  \left| \Gamma_{\kappa} \left( \mathcal{X} \right) (h) \right|^2$, for each $\mathcal{X} \in \mathbb{H}$, $\left\| \Gamma_{\kappa}^{k} \right\|_{\mathcal{L} \left( \mathbb{H} \right)} < 1$, for each $k \geq k_0$ and for some $k_0 \geq 1$ \citep[Lemma 1]{AlvarezLiebanaetal16}.
	\subsection{A two-steps specification test for the OU model}
	\label{sec:gof_OU}
	    As explained, the ARH(1) characterization of an OU process provided in~\eqref{arh_ou} will allow us to propose a two-stage methodology in this subsection with the aim of developing a specification test for these diffusion processes. In brief, the underlying idea will be, firstly, test whether a diffusion process $\left\lbrace \xi_t \right\rbrace_{t \in \mathbb{R}^{+}},$ splitted and characterized as $\left\lbrace \mathcal{X}_n (t) \coloneqq \xi_{nh+t},~t \in \left[0, h \right] \right\rbrace_{n \in \mathbb{Z}}$, constitutes an ARH(1) process (via Algorithm~\ref{al:ARHz}), i.e., we will test whether, for each $n \in \mathbb{Z}$,

    	\begin{equation}\label{null_two_stages}
        	\mathcal{H}_{0}^{(1)} \colon \mathcal{X}_n \text{ and } \mathcal{X}_{n-1} \text{ are linearly related via } \Gamma \in \mathcal{L}, \qquad \mathcal{X}_n \left( t \right) = \Gamma \left(\mathcal{X}_{n-1} \right) \left( t \right) + \mathcal{E}_n \left( t \right).
    	\end{equation}
    	
    	After the characterization of an ARH(1) process as a FLMFR model in~\eqref{eq_20}, the second stage will be to check the parametric form of the linear operator $\Gamma$, with the null hypothesis $\mathcal{H}_{0}^{(2)} \colon \Gamma \left( \mathcal{X} \right) (t) \coloneqq \Gamma_{\kappa} \left( \mathcal{X} \right) (t) = e^{-\kappa t} \mathcal{X} (h)$, for each $\mathcal{X} \in \mathbb{H} = L^2 \left([0,h], \mathcal{B}_{[0,h]}, \lambda + \delta_{(h)}\right)$ ($ \mathcal{X}_n (t) = \xi_{nh+t} = \xi_0 e^{-\kappa \left(nh + t \right)} + \sigma \int_{0}^{nh+t} e^{-\kappa (nh+t-s)} \dif W_s$, for each $n \in \mathbb{Z}$), via a functional F-test, as long as the linear null hypothesis $\mathcal{H}_{0}^{(1)}$ in~\eqref{null_two_stages} is not rejected, against the alternative of an unspecified FLMFR.
    	
    	As referred, a F-test will be implemented. In keeping with the classical F-statistic, the functional version proposed by \cite{ShenFaraway04} has been implemented, based on the residual sum of squared norm (RSSN) of functional errors \citep{Cuevasetal04}:
    	\begin{equation}\label{RSSN}
        	\text{RSSN}_{n} = \sum_{i=0}^{n-1} \left\| \mathcal{Y}_i - \hat{\mathcal{Y}}_i \right\|_{\mathbb{H}}^2, \qquad \mathcal{Y}_i = \Gamma \left( \mathcal{X}_i \right) + \mathcal{E}_i, \qquad F_n = \frac{\text{RSSN}_{n}^{\text{OU}} - \text{RSSN}_{n}^{\text{FLMFR}}}{\text{RSSN}_{n}^{\text{FLMFR}}}, 
    	\end{equation}
    	where $\text{RSSN}_{n}^{\text{OU}}$ denotes the RSSN of functional errors under the assumption that  the linear operator is parametric defined as $\Gamma \left( \mathcal{X} \right) (t) \coloneqq \Gamma_{\kappa} \left( \mathcal{X} \right) (t) = e^{-\kappa t} \mathcal{X} (h)$ (i.e., under null hypothesis $\mathcal{H}_{0}^{(2)}$), whereas $\text{RSSN}_{n}^{\text{FLMFR}}$ denotes the RSSN of functional errors of an unrestricted FLMFR, whose estimator is given by the FPCR-L1S estimator provided in Section~\ref{sec:flmfr}. In the former case, and estimator of $\Gamma_{\kappa} \left( \mathcal{X}\right) (t) = e^{-\kappa t} \mathcal{X} (h)$ is required, in terms of the maximum likelihood sample-dependent estimator $\hat{\kappa}_n = \frac{- \int_{0}^{nh} \xi_t \dif \xi_t}{\int_{0}^{nh} \xi_{t}^2 \dif t}$  proposed in \cite{AlvarezLiebanaetal16}, whose strong-consistency was therein proved. Henceforth, $\hat{\Gamma}_{\kappa} \left( \mathcal{X}\right) (t) \coloneqq \Gamma_{\hat{\kappa}_n} \left( \mathcal{X}\right) (t) = e^{-\hat{\kappa}_n t} \mathcal{X} (h)$ is considered. In Algorithm~\ref{al:testOU}, a summary of the testing procedure is presented, via Bonferroni correction to counteract the multiple tests.
        
        \medskip
        
        \begin{algorithm} \label{al:testOU}
    		Let $\left\lbrace \xi_t \right\rbrace_{t \in \left[0, T \right]}$ be a stochastic processes on a given filtered space $\left( \Omega, \left\lbrace \mathcal{A}_t \right\rbrace, \mathbb{P} \right)$ where $\xi_t$ is $\mathcal{A}_t$-measurable, for each $t \in \left[0, T \right]$, given by a parametric time-homogeneous SDE (see \eqref{homogeneous_SDE}), under \textbf{Assumptions~\ref{as:41}--\ref{as:42}} and null long term mean. We proceed as follows:
    		
    		\textbf{Stage 1 ($\mathcal{H}_{0}^{(1)}$):} testing if $\left\lbrace \xi_t \right\rbrace_{t \in \left[0, T \right]}$ can be reinterpreted as a functional ARH(1) model.
    		\begin{enumerate}
    			\item Split $\left\lbrace \xi_t \right\rbrace_{t \in \left[0, T \right]}$ into a set of functional paths $\left\lbrace \mathcal{X}_i (t) \coloneqq \xi_{ih+t},~t \in \left[0, h \right] \right\rbrace_{i=0}^{n-1}$ valued in $n$ subintervals, where $T = nh$ and $\mathcal{X}_i \in L^2 \left( [0, h], \mathcal{B}_{[0,h]}, \lambda + \delta_{(h)} \right)$.
    			\item Construct frv's $\widetilde{\mathcal{X}}_i (s)\coloneqq  \mathcal{X}_{i-1} (s)$,  $\widetilde{\mathcal{Y}}_i (t)\coloneqq \mathcal{X}_{i}(t)$ and $\widetilde{\mathcal{E}}_i (t)\coloneqq \mathcal{E}_{i}(t)$, all of them centered, with $\widetilde{\mathcal{Y}}_i = \widetilde{\Gamma} \left(\widetilde{\mathcal{X}}_i \right) + \widetilde{\mathcal{E}}_i$, for any $i=1,\ldots,n-1$, and $s,t \in [0,h]$.
    			\item Obtain the FPC of $\left\lbrace \widetilde{\mathcal{X}}_i  \right\rbrace_{i=1}^{n-1}$ and $\left\lbrace \widetilde{\mathcal{Y}}_i  \right\rbrace_{i=1}^{n-1}$, choose initial $(p,q)$ required for $\text{EV}_p$ = $\text{EV}_q$ = 0.99, and achieve their $p$- and $q$-truncated FPC scores $\widetilde{\mathbf{X}}_p$  and $\widetilde{\mathbf{Y}}_q$, respectively. We then compute the FPCR-L1S estimator $\widehat{\widetilde{\mathbf{B}}}_{\widetilde{p},q}^{(\lambda), C}$ and the associated residuals. As detailed in \textbf{Algorithm~\ref{al:ARHz}}, we compute the statistic $\text{PCvM}_{n,\widetilde{p},q}$ in \eqref{eq_18}, the bootstrapped errors $\left\lbrace \mathbf{e}_{i,q}^{\ast b} \right\rbrace_{i=1,\ldots,n-1}^{b=1,\ldots,B}$, whit $B$ replicates, and the bootstrapped statistic $\text{PCvM}_{n,\widetilde{p},q}^{\ast b}$.
    			\item Estimate the p-value as $p^{(1)} \coloneqq \#\left\lbrace \text{PCvM}_{n,\widetilde{p},q} \leq  \text{PCvM}_{n,\widetilde{p},q}^{\ast b}\right\rbrace / B$. If $\mathcal{H}_{0}^{(1)}$ in~\eqref{null_two_stages} is rejected, the procedure stops. Otherwise, go to stage 2.
    		\end{enumerate}
    		\textbf{Stage 2 ($\mathcal{H}_{0}^{(2)}$):} testing if $\left\lbrace \xi_t \right\rbrace_{t \in \left[0, T \right]}$, characterized as an ARH(1) model (under linearity), constitutes an OU process via F-test, against the alternative of an unspecified FLMFR.
    		\begin{enumerate}
    			\setcounter{enumi}{4}
    			\item If $\mathcal{H}_{0}^{(1)}$ in~\eqref{null_two_stages} is not rejected, calculate the F-statistic as $F_n = \frac{\text{RSSN}_{n}^{\text{OU}} - \text{RSSN}_{n}^{\text{FLMFR}}}{\text{RSSN}_{n}^{\text{FLMFR}}}$, where $\text{RSSN}_{n}$ is computed as~\eqref{RSSN}, with  $\text{RSSN}_{n}^{\text{OU}} = \sum_{i=1}^{n-1} \left\| \widetilde{\mathcal{Y}}_{i}(t) - e^{-\hat{\kappa}_n t} \widetilde{\mathcal{X}}_{i}(h)  \right\|_{\mathbb{H}}^{2}$.
    			\item We implement a parametric bootstrap, by simulating a set of OU processes $\left\lbrace \xi_{t}^{\ast b} \right\rbrace_{t \in [0,T]}^{b=1,\ldots,B}$, driven by $\dif \xi_{t}^{\ast b} = - \hat{\kappa}_n \xi_{t}^{\ast b} \dif t + \hat{\sigma}_n \dif W_{t}^{\ast b}$, for each $b = 1,\ldots,B$, where $B$ is the number of replicates, $\hat{\sigma}_n = \displaystyle \argmin_{\sigma_0} \frac{1}{N-1} \sum_{i=0}^{N-1} \left(\ln \left(\sigma_{0}^{2} \right) + \frac{\left( \xi_{t_{i+1}} - \xi_{t_{i}} \right)^2}{\sigma_{0}^{2} \Delta} \right)$ is the consistent estimator of $\sigma$ (\citealp{CorradiWhite99}) and $\hat{\kappa}_n$ is the referred estimator of $\kappa$, where $\left\lbrace t_0, \ldots, t_N \right\rbrace$ are the grid points of $[0,T]$, with $T = nh$, and $\Delta$  the discretization step.
    			\item With the bootstrapped $\left\lbrace \xi_{t}^{\ast b} \right\rbrace_{t \in [0,T]}^{b=1,\ldots,B}$, we recompute $F_{n}^{\ast b} = \frac{\text{RSSN}_{n}^{OU,\ast b} - \text{RSSN}_{n}^{FLMFR,\ast b}}{\text{RSSN}_{n}^{FLMFR,\ast b}}$, such that an estimator $\hat{\kappa}_{n}^{\ast b}$ is required to be recomputed, as well as a FPCR-L1S estimator for $\Gamma_{\hat{\kappa}_{n}}$, for each $b=1,\ldots,B$.
    			\item Estimate the p-value as $p^{(2)} \coloneqq \#\left\lbrace F_{n} \leq  F_{n}^{\ast b}\right\rbrace / B$.
    		\end{enumerate}
    		\textbf{Bonferroni correction: stage 1 + stage 2}. As we are conducting multiple tests, given the corresponding $p$-values $p^{(1)}$ and $p^{(2)}$ for testing the hypotheses $\mathcal{H}_{0}^{(1)}$ and $\mathcal{H}_{0}^{(2)}$, the Bonferroni correction is used to set an upper bound on the significance level $\alpha$ \citep{Miller81}, by rejecting  $\mathcal{H}_0 = \lbrace \mathcal{H}_{0}^{(1)}, \mathcal{H}_{0}^{(2)} \rbrace$ if any p-value $p^{(1)}$ or $p^{(2)}$ is less than $\alpha/2$. As referred, the second stage in \textbf{Algorithm~\ref{al:testOU}} is omitted if $\mathcal{H}_{0}^{(1)}$ is rejected, respect to $\alpha/2$.
	    \end{algorithm}

	\subsection{Simulation study}
	\label{sec:simulations}
	
    	The finite sample properties of the specification test is illustrated with $1\,000$ Monte Carlo replicates and $B = 1\,000$ bootstrap resamples, with $n \in \left\lbrace 50, 150, 250, 350, 500, 1000, 5000\right\rbrace$, such that trajectories $\left\lbrace \mathcal{X}_i (t) \coloneqq \xi_{ih + t} \right\rbrace_{i=0}^{n-1}$ are valued in 101 equispaced points in $[0,1]$. The FPCR-L1S estimator was used with $\text{EV}_p = \text{EV}_q = 0.995$. As reflected in Table~\ref{tab:2step_hyp}, several scenarios have been simulated. Under $\mathcal{H}_0$, we simulate  sample paths for the centered OU model in~\eqref{integral_OU}. On the other hand, concerning the power, we simulate different interest rate models, as the CKLS model  \citep{chan1992empirical}, the Inverse Feller (IF) model  \citep{AhnGao99} and the Aït-Sahalia (AS) model \citep{AitSahalia96}. Similar parameters of the CKLS model to those ones proposed in \cite{HongLi04},  to examine different persistent dependence and  volatility, were tested. The IF and AS models are given by $\dif \xi_t = \xi_t (\kappa - (\sigma^2 - \kappa \mu)\xi_t) \dif t + \sigma \xi_t^{3/2} \dif W_t$ and $\dif \xi_t = (\tau_{-1}\xi_t^{-1} + \tau_0 + \tau_1 \xi_t + \tau_2 \xi_t^2 ) \dif t + \sigma \xi_t^{3/2} \dif W_t$, with similar set of parameters as proposed in \cite{AhnGao99} and \cite{ait2001transition}, respectively  (see Table~\ref{tab:2step_hyp}). For testing deviations from the null hypothesis, we also consider the radial OU process\footnote{The drift function is non-Lipschitz and unbounded, however, there exist a unique weak SDE solution.}, given by $\dif \xi_t = (\lambda \xi_t^{-1} - \kappa \xi_t) \dif t + \sigma \dif W_t$, which corresponds with the OU if $\lambda = 0$. A null model has been also tested, given by $\dif \xi_t = \sigma \dif W_t$, with $\sigma \in \lbrace 0.1, 0.5 \rbrace$.

    	\begin{table}[t]
    	\centering
    	\footnotesize
    	\renewcommand{\arraystretch}{0.6}
    	\caption{Summary of simulated scenarios.}
    	\begin{tabular}{lcll}
    		\toprule 
    		Notation &  Description & \multicolumn{1}{c}{Model} & Parameters and scenarios  \\
    		\midrule
    		\multirow{2}{*}{$\mathcal{H}_{0}$} & OU & \multirow{2}{*}{$\dif \xi_t = -\kappa \xi_t \dif t + \sigma \dif W_t$} & $\kappa \in \lbrace 0.2,0.5,0.8 \rbrace$ \\
    		& (centered) & & $\sigma^2 \in \lbrace 0.008,0.05,0.15,0.50,0.75 \rbrace$ \\
    		\hline 
    		\multirow{2}{*}{$\mathcal{H}_{1, Null}$} & \multirow{2}{*}{Null} & \multirow{2}{*}{$\dif \xi_t = \sigma \dif W_t$} & S1: $\sigma = 0.1$ \\
    		&  & &  S2: $\sigma = 0.5$ \\
    		\hline
    		\multirow{2}{*}{$\mathcal{H}_{1, IF}$} & \multirow{2}{*}{Inverse-Feller} & $\dif \xi_t = \xi_t (\kappa - (\sigma^2 - \kappa \mu)\xi_t) \dif t + $ & 
    		\multirow{2}{*}{$(\kappa, \mu, \sigma^2) = (0.364, 0.08, 1.6384)$} \\
    		 &  & $\phantom{\dif \xi_t = }~\sigma \xi_t^{3/2} \dif W_t$ &  \\
    		\hline
    		\multirow{2}{*}{$\mathcal{H}_{1, AS}$}     & \multirow{2}{*}{Aït-Sahalia}    & $\dif \xi_t = (\tau_{-1}\xi_t^{-1} + \tau_0 + \tau_1 \xi_t + $ & 
    		$(\tau_{-1}, \tau_0, \tau_1, \tau_2, \sigma) = (0.00107,-0.0517 $ \\
    		 &  & $\phantom{\dif \xi_t = }~\tau_2 \xi_t^2 ) \dif t + \sigma \xi_t^{3/2} \dif W_t$ & 
    		$\phantom{(\alpha_{-1}, \alpha_0, \alpha_1, \alpha_2) = (}~0.877, -4.604,0.8)$ 
    		\\
    		\hline
    		\multirow{3}{*}{$\mathcal{H}_{1, CKLS}$} & \multirow{3}{*}{CKLS} & \multirow{3}{*}{$\dif \xi_t = \kappa(\mu - \xi_t) \dif t + \sigma \xi_t^{\gamma} \dif W_t$} & S1: $(\mu, \kappa, \sigma, \gamma) = (0.09, 0.9, 0.5, 1.5)$ \\
    		&  & &  S2: $(\mu, \kappa, \sigma, \gamma) = (0.09, 0.2, 1.5, 1.5)$ \\
    		&  & &  S3: $(\mu, \kappa, \sigma, \gamma) = (0.09, 0.2, 3, 1.5)$ \\
    		\hline
    		\multirow{4}{*}{$\mathcal{H}_{1, ROU}$} & \multirow{4}{*}{Radial OU} & \multirow{4}{*}{$\dif \xi_t = (\lambda \xi_t^{-1} - \kappa \xi_t) \dif t + \sigma \dif W_t$} & S1: $(\lambda, \kappa, \sigma) = (0.05, 0.1, 0.5)$ \\
    		&  & &  S2: $(\lambda, \kappa, \sigma) = (0.075, 0.1, 0.5)$ \\
    		&  & &  S3: $(\lambda, \kappa, \sigma) = (0.1, 0.1, 0.5)$ \\
    		&  & &  S4: $(\lambda, \kappa, \sigma) = (0.125, 0.1, 0.5)$ \\
    		\bottomrule
    	\end{tabular}%
    	\label{tab:2step_hyp}
    	\end{table}
    	Tables~\ref{tab:2step_sizes}--\ref{tab:2step_power} display the empirical sizes and powers, respectively, with $\alpha = 0.05$ as nominal level. In the former table, regarding the calibration of test, the empirical sizes are close to $\alpha$ through all values of $\sigma$ and $\kappa$. As expected, scenarios with $n=50$ exhibit over-rejection for larger $\kappa$ values, owing to $\kappa$ represents the speed of reversion at which trajectories are rearranged around $\mu$, and then, greater values of $\kappa$ lead to more similar paths to be discriminated, since $\displaystyle \lim_{t \to \infty} {\rm Var } \left[\xi_t \right] = \frac{\sigma^{2}}{2\kappa}$ constitutes the long term variance. This behavior is overcome as $n$ increases: just 2 of the 90 scenarios (2.22\% of cases) with $n \in \left\lbrace 150, 250, 350, 500, 1000, 5000 \right\rbrace$ display slightly over-rejections (boldfaced rejection rates). Note that sample size in the functional data framework refers to the number of curves, the actual number of observations in the simulation reaches $500\,000$ for the $n = 5\,000$ scenario, as the curves are evaluated in $101$ equispaced points.

        \begin{table}[t]
        \caption{(a) Empirical rejection rates under the null hypothesis. The rejection rates are boldfaced if they lie outside a 95\%-confidence interval for $\alpha = 0.05$. (b) Empirical rejection rates under the alternative hypothesis, from scenarios displayed in Table~\ref{tab:2step_hyp}.}
        \begin{adjustwidth}{-1.0cm}{}
        \begin{subtable}{0.45\linewidth}
          \centering
    		\footnotesize
    		\renewcommand{\arraystretch}{0.8}
    		\caption{\label{tab:2step_sizes} Size simulation.}
    		\begin{tabular}{c@{\hspace{0.5em}}c@{\hspace{0.6em}}c@{\hspace{0.5em}}c@{\hspace{0.5em}}c@{\hspace{0.5em}}c@{\hspace{0.5em}}c@{\hspace{0.5em}}c@{\hspace{0.5em}}c}
    			\toprule
    			& &  \multicolumn{7}{c}{$n$}  \\ 
    			\cmidrule{3-9} 
    			$\sigma^2$ & $\kappa$ & 50 & 150 & 250 & 350 & 500 & 1000 & 5000 \\ 
    			\midrule
    			\multirow{3}{*}{0.008} & 0.2 & 0.048 & 0.043 & 0.049 & 0.053 & 0.045 & 0.051 & 0.058 \\
    			                       & 0.5 & 0.047 & 0.053 & 0.048 & 0.046 & 0.050 & 0.049 & 0.052 \\
    			                       & 0.8 & \textbf{0.078} & 0.048 & 0.051 & 0.056 & 0.042 & 0.058 & 0.054 \\
    			\hline 
    			\multirow{3}{*}{0.05}  & 0.2 & 0.043 & 0.051 & 0.054 & 0.048 & 0.053 & 0.052 & 0.061 \\
    			                       & 0.5 & 0.051 & 0.057 & 0.055 & 0.055 & 0.061 & 0.048 & 0.054 \\
    			                       & 0.8 & \textbf{0.102} & 0.060 & 0.052 & \textbf{0.075} & 0.057 & 0.053 & 0.044 \\
    			\hline 
    			\multirow{3}{*}{0.15}  & 0.2 & 0.050 & 0.045 & 0.048 & 0.053 & 0.046 & 0.051 & 0.060 \\
    			                       & 0.5 & 0.045 & 0.053 & 0.053 & 0.046 & 0.050 & 0.050 & 0.048 \\
    			                       & 0.8 & \textbf{0.078} & 0.051 & 0.052 & 0.056 & 0.043 & 0.056 & 0.056 \\
    			\hline 
    			\multirow{3}{*}{0.50}  & 0.2 & 0.051 & 0.046 & 0.053 & 0.046 & 0.051 & 0.052 & 0.058 \\
    			                       & 0.5 & 0.050 & 0.056 & 0.059 & 0.054 & 0.053 & 0.051 & 0.061 \\
    			                       & 0.8 & \textbf{0.096} & \textbf{0.065} & 0.057 & 0.059 & 0.056 & 0.059 & 0.053 \\
    			\hline 
    			\multirow{3}{*}{0.75}  & 0.2 & 0.046 & 0.044 & 0.049 & 0.052 & 0.047 & 0.051 & 0.053 \\
    			                       & 0.5 & 0.048 & 0.054 & 0.050 & 0.047 & 0.051 & 0.048 & 0.052 \\
    			                       & 0.8 & \textbf{0.080} & 0.058 & 0.053 & 0.054 & 0.041 & 0.057 & 0.050 \\ 
    			\bottomrule
    		\end{tabular}
        \end{subtable}%
        \hspace{1.4cm}
        \begin{subtable}{0.45\linewidth}
          \footnotesize
    		\centering
    			\renewcommand{\arraystretch}{1.05}
    		\caption{\label{tab:2step_power} Power simulation.}
    		\begin{tabular}{l@{\hspace{0.5em}}c@{\hspace{0.5em}}c@{\hspace{0.5em}}c@{\hspace{0.5em}}c@{\hspace{0.5em}}c@{\hspace{0.5em}}c@{\hspace{0.5em}}c}
    			\toprule 
    			 & \multicolumn{7}{c}{$n$} \\
    			 \cmidrule{2-8} 
    			Scenario &  50    & 150   & 250   & 350   & 500 & 1000 & 5000 \\
    			\midrule 
    			$\mathcal{H}_{1, ROU}^{S1}$ & 0.043 & 0.145 & 0.241 & 0.340 & 0.481 & 0.818 & 0.945 \\
    			$\mathcal{H}_{1, ROU}^{S2}$ & 0.060 & 0.232 & 0.390 & 0.530 & 0.744 & 0.956 & 0.989 \\
    			$\mathcal{H}_{1, ROU}^{S3}$ & 0.070 & 0.280 & 0.487 & 0.722 & 0.900 & 0.976 & 1.000 \\
    			$\mathcal{H}_{1, ROU}^{S4}$ & 0.065 & 0.276 & 0.586 & 0.807 & 0.945 & 0.986 & 1.000 \\
    			\hline 
    			$\mathcal{H}_{1, Null}^{S1}$ & 0.097 & 0.322 & 0.583 & 0.741 & 0.907 & 1.000 & 1.000 \\
    			$\mathcal{H}_{1, Null}^{S2}$ & 0.108 & 0.317 & 0.566 & 0.765 & 0.919 & 1.000 & 1.000 \\
    			\hline 
    			$\mathcal{H}_{1, IF}$ & 0.455 & 0.750 & 0.836 & 0.870 & 0.928 & 0.976 & 1.000 \\
    			\hline 
    			$\mathcal{H}_{1, AS}$ & 0.139 & 0.370 & 0.499 & 0.611 & 0.756 & 0.958 & 1.000 \\
    			\hline 
    			$\mathcal{H}_{1, CKLS}^{S1}$ & 0.179 & 0.277 & 0.360 & 0.488 & 0.667 & 0.763 & 0.886 \\
    			$\mathcal{H}_{1, CKLS}^{S2}$ & 0.319 & 0.635 & 0.727 & 0.738 & 0.811 & 0.895 & 0.994 \\
    			$\mathcal{H}_{1, CKLS}^{S3}$ & 0.690 & 0.933 & 0.985 & 0.993 & 1.000 & 1.000 & 1.000 \\
    			\bottomrule 
    		\end{tabular}
        \end{subtable}%
        \end{adjustwidth}
    \end{table}

	With respect to empirical power (Table~\ref{tab:2step_power}), deviations from the null, by augmenting  $\lambda$ in the radial OU, show the increasing power of the test as $n$ increases. The null model also exhibits high empirical powers, just like the nonlinear drift alternatives, specially the IF model, displaying great empirical powers even for small sample sizes. Concerning the CKLS model, all the scenarios provide increasing rejection rates, such that the higher the volatility parameter, the more empirical power. A slightly poor performance is obtained for the more complex AS model, although  empirical powers seem to tend to one as $n$ increases.
	\section{Real data applications}
	\label{sec:real_data}
	
    	Motivated by the extensive use of diffusion processes in finance, commonly used to model currency exchange rates \citep{ball1994target} and intra-day patterns in the foreign exchange market \citep{AndersenBollerslev97}, we now illustrate a real-data application of our OU specification test, in the context of high-frequency financial data. Unlike univariate frameworks, where a vast time window is required for getting properly sample sizes, and thus, certain properties essential for long horizon asymptotics (e.g., ergodicity) would not be achieved, we consider a finite time observation window, where the functional data scheme allows to capture the dynamic of the process \citep{Mulleretal11}. 
    	
\begin{figure}[t]
        	\centering
        	\includegraphics[width=1\linewidth]{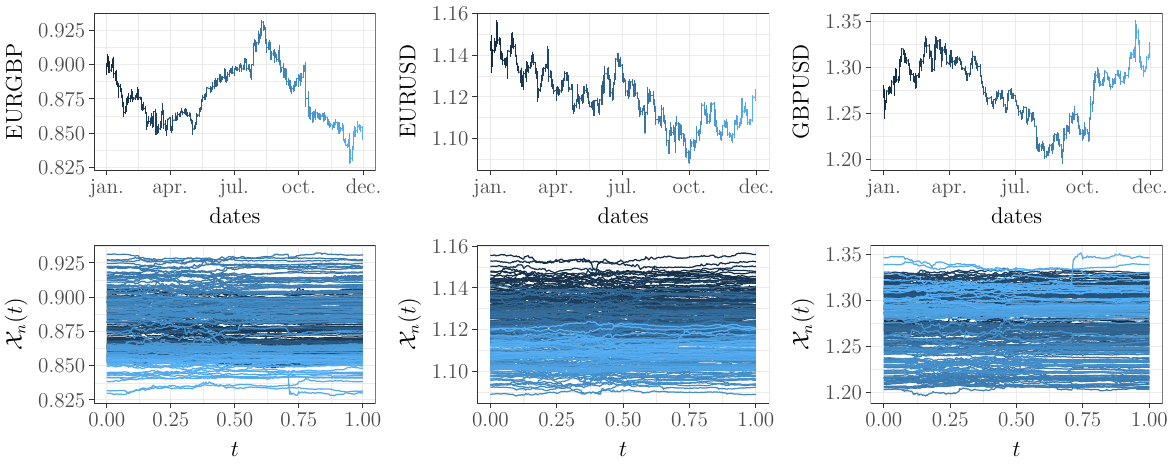}
        	\caption{From left to right: EURGBP, EURUSD and GBPUSD exchange rates throughout 2019. On the top, the observed stochastic process; at the bottom, daily splitted paths.}
        	\label{fig:sde}
        \end{figure}

    	The three datasets considered consist on currency pair rates Euro-British pound (EURGBP), Euro-US Dollar (EURUSD), British pound-US Dollar (GBPUSD), determined in the foreign exchange market, such that data was recorded every 5 minutes from January 1, 2019 to December 31, 2019.
    	Figure~\ref{fig:sde} shows the exchange rates (top)  with $73\,440$ observations, and the splitted paths (bottom) $\left\lbrace \mathcal{X}_i (t) \right\rbrace_{i=1}^n$ featuring $n=255$ daily curves. The daily curves are valued in $\mathbb{H} = L^2 \left([0,1], \mathcal{B}_{[0,1]}, \lambda + \delta_{(1)} \right)$, where the interval $[0,1]$ accounts for a 1-day window,  discretized in $288$ equispaced grid points. Table~\ref{tab:pval} shows the empirical $p$-values for the three datasets, with $B = 1\,000$ bootstrap resamples, at each of the two-stages defined in \textbf{Algorithm~\ref{al:testOU}}. In this way, we test at stage one the null hypothesis that the daily curves $\left\lbrace \mathcal{X}_i (t) \right\rbrace_{i=1}^n$ constitute an ARH(1) process,
    	\begin{equation*}
    	\mathcal{H}_{0}^{(1)} \colon \mathcal{X}_n \text{ and } \mathcal{X}_{n-1} \text{ are linearly related via } \Gamma \in \mathcal{L},
    	\end{equation*}
    	with $\mathcal{X}_n \left( t \right) = \Gamma \left(\mathcal{X}_{n-1} \right) \left( t \right) + \mathcal{E}_n \left( t \right)$. At stage two, a F-test is implemented to test the parametric form of the OU process, $\mathcal{H}_{0}^{(2)} \colon \Gamma \left( \mathcal{X} \right) (t) \coloneqq \Gamma_{\kappa} \left( \mathcal{X} \right) (t) = e^{-\kappa t} \mathcal{X} (h)$.
    	%
        
    	Regarding the $p$-values from Table~\ref{tab:pval}, the null hypothesis of OU process in the EURGBP series is rejected, at a 5\% level, since it does not seem to follow an stationary ARH(1) process (stage one) when the annual trajectory is considered as daily curves. However, it is not rejected for the EURUSD and GBPUSD exchange rates, where a simple model with constant volatility function as the OU seems to capture the dependence of the series. The nature of the sampling mechanism could generate a different result, as classical data analysis for diffusion processes deals with monthly, weekly or daily data, at most. As the sampling frequency is not dictated by the data, usually a large fraction of data is discarded, not without loss of information. In our data example, sampling daily would mean to retain only 255 observations, from a total of $73\,440$, working with one observation a day instead of the whole daily trajectory. Therefore, when dealing with intra-day data, conducting the study using a functional framework takes advantage of the information retained in the shape of the curve, treating the daily curves as a statistical object, instead of a collection of individual observations.

        \begin{table}[H]
		\caption{$p$-values under the null hypothesis that series follows a centered OU process.}
		\centering
		\renewcommand{\arraystretch}{0.6}
		{\small
		\begin{tabular}{lccc}
			\toprule
			Two-stages test & EURGBP & EURUSD & GBPUSD \\
			\midrule 
			\textbf{Stage 1}: ARH(1) GoF test & 0.020 & 0.339 & 0.091 \\
			\textbf{Stage 2}: F-test          & 0.349 & 0.253 & 0.510 \\  
			\bottomrule 
		\end{tabular} 
		}
		\label{tab:pval}
	    \end{table}

	\section{Conclusions and open research lines}
	\label{sec:conc}
	
	In this paper we have proposed a novel GoF test for ARH(z) models, against an unspecified alternative, based on the their characterization as FLMFR. No competitors, against an unrestricted alternative, could be found, but empirical results under different alternative were compared with \cite{KokoszkaReimherr12} (\text{KR}) on determining the order $z$. Outcomes showed that our proposal is well calibrated under the null hypothesis and exhibited a great power through several alternative scenarios, being competitive even against nonlinear alternatives where the \text{KR} test cannot reach a suitable power. 
	
	On the other hand, a novel specification test for the Vasicek model was proposed, firstly characterizing it as an ARH(1) model, and secondly, testing the parametric form, under linearity, via functional F-test, calibrated by a parametric bootstrap. The finite sample behaviour was also illustrated, generating the data under parametric families of diffusion processes, providing the empirical evidence that the two-stages specification test is well-calibrated, and detected different alternative models widely used in finance. 
	As traditional data analysis ignores the information contained in the shape of the daily curves, the functional data approach could be more suitable when working with intra-day or high-frequency data, as it can extract the additional information of the curves. The classical diffusion process framework regards the whole observation sequence as one sample path, ignoring the pattern of the process across days. Working in a functional data setting, we take the advantage of the information retained in the shape of the curve, treating the daily curves as a statistical object, instead of a collection of individual observations. Our test is illustrated with an application to daily currency exchange rates curves, with 5-minute intervals, where we found that the Vasicek model does capture the underlying dynamic changes of the daily curves for the EURUSD and GBPUSD pairs.

	Future applications could be developed, since the referred specification test may be extended to other diffusion processes as long as they may be characterized as ARH($z$) models. The tests here derived may be also extended to Hilbertian moving-average processes and ARH processes with exogenous variables. 

    \bigskip
    
    \begin{center}
    {\large\bf SUPPLEMENTARY MATERIAL}
    \end{center}
    
    \begin{description}
        \item[Supplementary Material (pdf):] Document containing fruitful discussions on theoretical aspects in functional data analysis, main details on estimating and testing FLMFR and ARH($z$) models, and the proofs of main  theoretical results.
        \item[Software:] A companion R software,  allowing to replicate all tests and applications, is freely available at \url{github.com/dadosdelaplace/gof-test-arh-ou-process}
    \end{description}

	\bibliographystyle{plainnat}
	\bibliography{Biblio}

\end{document}